%% file: ms-emulapj.tex
\documentclass[12pt,preprint]{emulateapj}
\journalinfo{}
\submitted{To appear in the Astrophysical Journal}
\usepackage{natbib}

\newcommand{\fxunits}{\mbox{ergs cm$^{-2}$ s$^{-1}$}}
\newcommand{\lxunits}{\mbox{ergs s$^{-1}$}}

\def\hpone{h$_{\rm 50}$ }

\begin{document}

\title{Evolution of the Cluster X-ray Luminosity Function}

\author{C.R. Mullis\altaffilmark{1},
        A. Vikhlinin\altaffilmark{2,3},
        J.P. Henry\altaffilmark{4},
        W. Forman\altaffilmark{2},
	I.M. Gioia\altaffilmark{5},\\
        A. Hornstrup\altaffilmark{6},
        C. Jones\altaffilmark{2},
        B.R. McNamara\altaffilmark{7}, and
        H. Quintana\altaffilmark{8}}

\altaffiltext{1}{European Southern Observatory, Headquarters,
                 Karl-Schwarzschild-Strasse 2, Garching bei M\"unchen D-85748,
                  Germany, cmullis@eso.org}
\altaffiltext{2}{Harvard-Smithsonian Center for Astrophysics, 
                 60 Garden Street, Cambridge, MA 02138, USA}
\altaffiltext{3}{Space Research Institute, Profsoyuznaya 84/32, Moscow, Russia}
\altaffiltext{4}{Institute for Astronomy, University of Hawai`i, 
		 2680 Woodlawn Drive, Honolulu, HI 96822, USA}
\altaffiltext{5}{Istituto di Radioastronomia del CNR-INAF, via Gobetti 101, 
                 Bologna, I-40129, Italy}
\altaffiltext{6}{Danish Space Research Institute, Juliane Maries Vej 30, 
		 Copenhagen 0, DK-2100, Denmark}
\altaffiltext{7}{Department of Physics and Astronomy, Ohio University, 
                 Athens, OH 45701, USA}
\altaffiltext{8}{Departamento de Astronomia y Astrofisica, 
                 Pontificia Universidad Catolica de Chile, Casilla 306, 
                 Santiago, 22, Chile}

\shorttitle{CLUSTER XLF EVOLUTION} \shortauthors{MULLIS ET AL.}

\begin{abstract}

We report measurements of the cluster X-ray luminosity function out to
$z=0.8$ based on the final sample of 201 galaxy systems from the 160
Square Degree {\em ROSAT\/} Cluster Survey.  There is little evidence for any
measurable change in cluster abundance out to $z\sim 0.6$ at
luminosities less than a few times \mbox{10$^{44}$ $h_{50}^{-2}$}
\lxunits~\mbox{(0.5--2.0~keV)}.\@ However, between $0.6 < z < 0.8$ and
at luminosities above \mbox{10$^{44}$ $h_{50}^{-2}$} \lxunits, the
observed volume densities are significantly lower than those of the
present-day population.  We quantify this cluster deficit using
integrated number counts and a maximum-likelihood analysis of the
observed luminosity-redshift distribution fit with a model luminosity
function.  The negative evolution signal is $>3\sigma$ regardless of
the adopted local luminosity function or cosmological framework.  Our
results and those from several other surveys independently confirm the
presence of evolution.  Whereas the bulk of the cluster population
does not evolve, the most luminous and presumably most massive
structures evolve appreciably between $z=0.8$ and the present.
Interpreted in the context of hierarchical structure formation, we are
probing sufficiently large mass aggregations at sufficiently early
times in cosmological history where the Universe has yet to assemble
these clusters to present-day volume densities.

\end{abstract}


%----------------- SECTION 1: INTRODUCTION -------------------------------
\section{Introduction} 
\label{Introduction} 

Galaxy clusters form via the gravitational amplification of rare, high
peaks in the cosmic matter density field.  The redshift evolution of
cluster abundance depends on the growth rate of density perturbations
which is, in turn, sensitive to the mean cosmic matter density
($\Omega_{M}$) and, to a lesser extent, the dark-energy density
($\Omega_{\Lambda}$).  Thus observations of cluster space density
with sufficient temporal sampling can provide powerful cosmological
constraints \citep[e.g.,][]{Oukbir1992,Eke1998,Bahcall1999}.

A directly measurable and robust diagnostic of cluster abundance is
the X-ray luminosity function (XLF), that is the volume density of
clusters per luminosity interval.  \mbox{X-ray} selected galaxy
clusters are particularly well-suited for this type of
analysis. Clusters are efficiently detected at \mbox{X-ray}
wavelengths to high redshift (currently out to $z \sim 1.25$) thus
providing the leverage for evolutionary studies \citep*[see review
by][]{Rosati2002b}.  The resulting samples feature high statistical
completeness which is clearly important for deriving reliable number
counts.  Since X-ray surveys have well-defined selection functions, it
is straight-forward to convert these number counts into
volume-normalized measures such as the luminosity function.  Finally,
given the strong correlation between X-ray luminosity and cluster
mass, it is possible to transform the observed XLF into the cluster
mass function which is the fundamental relation in the theoretical
treatment.

Though the cluster XLF was first measured with an \mbox{X-ray}
flux-limited sample over two decades ago \citep{Piccinotti1982}, a
definitive characterization of its evolution has proven very
difficult.  The latter is a particularly important test in
observational cosmology because the variation of the cluster XLF as a
function of redshift reflects the evolution of the cluster mass function.
Such a measure allows for strong constraints on $\Omega_{M}$ even
accounting for uncertainty in the luminosity-mass relation
\citep[e.g.,][]{Borgani2001b}.  Early theoretical predictions
\citep[e.g.,][]{Kaiser1986} postulated dramatic positive evolution in
the XLF where the volume density of clusters of a fixed luminosity
would increase with redshift.  This would be an ``observer-friendly''
universe since the loss in sensitivity at high redshift in a
flux-limited survey would be offset by the growing population of
detectable sources.  Contrary to this scenario, observations of the
cluster XLF range from zero evolution to {\em negative} evolution
depending on the redshifts and luminosities probed.  These findings
are consistent with current predictions for a low-density universe
where mild negative evolution is restricted to the most luminous,
high-redshift clusters while there is little change in the bulk of the
population \citep[e.g.,][and references therein]{Borgani2001}.

%................................................................
% FIGURE 1 --- IDL: /figs/fig1-lxzall.pro ---
\begin{figure*}
\epsscale{1}
\plotone{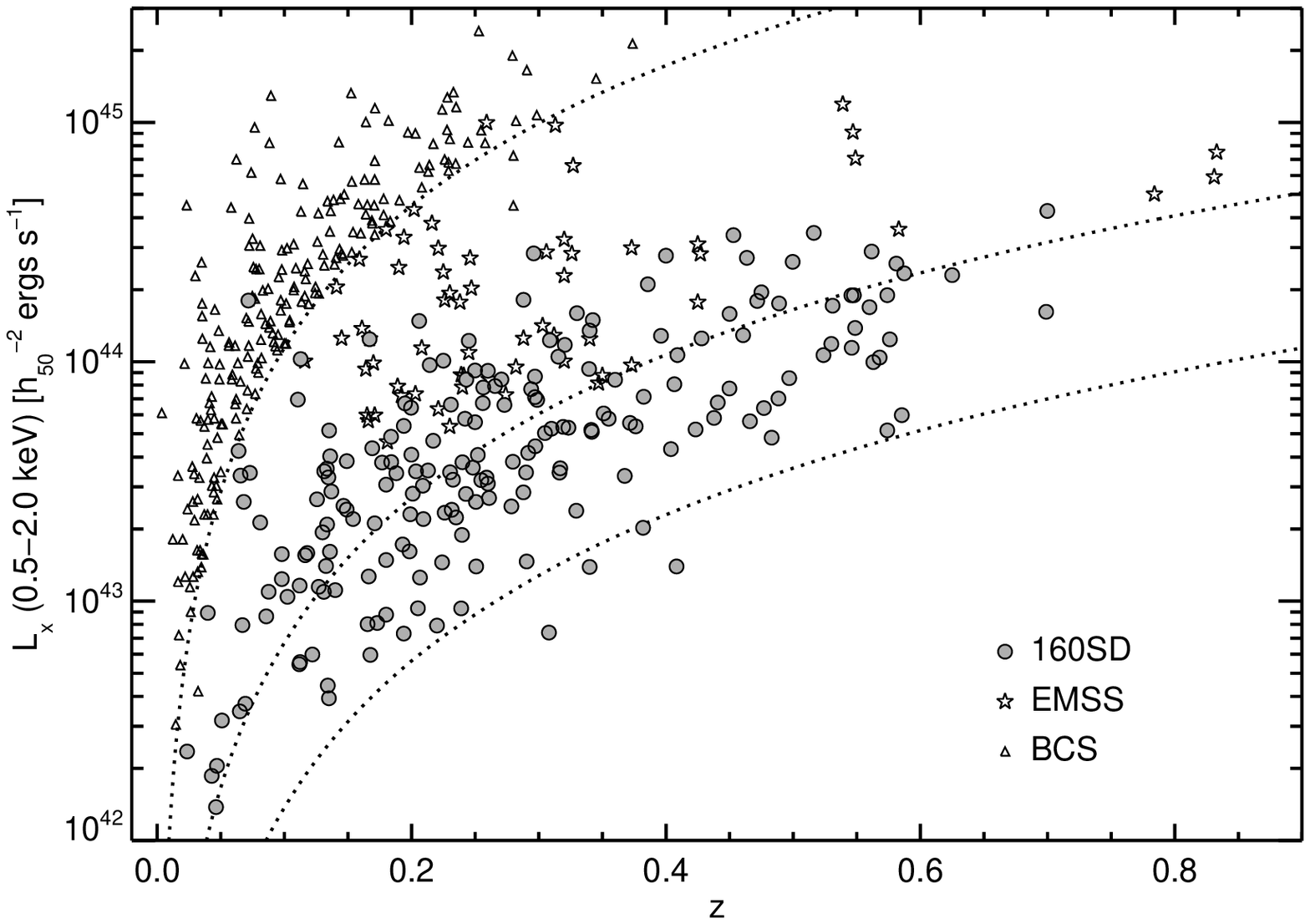}

\caption{X-ray luminosity and redshift distribution of the 160SD
\citep{Mullis2003a}, the EMSS (\citealt{Gioia1994} with updates from the
literature, e.g., \citealt{Henry2000}), and the BCS \citep{Ebeling1998}
cluster samples.  The dotted curves (left to right) are indicative
flux limits of \mbox{2.7 $\times$ 10$^{-12}$}, \mbox{1.5 $\times$
10$^{-13}$}, and \mbox{3 $\times$ 10$^{-14}$} \fxunits
\,\mbox{(0.5--2.0\,keV)}. The EMSS and BCS luminosities were converted
to this energy band assuming a Raymond-Smith plasma spectrum
\citep{Raymond1977} with a metallicity of 0.3 solar and the reported
gas temperature (either directly measured or estimated from the
luminosity-temperature relation). \label{fig:lxzall}}

\end{figure*}
%................................................................

Taking advantage of the first time an X-ray survey extended into
cosmologically interesting redshifts, \citet{Gioia1990a} and
\citet{Henry1992} used 67 clusters ($0.14 < z < 0.60$) from the
pioneering {\em Einstein} Extended Medium Sensitivity Survey
\citep[EMSS;][]{Gioia1990b,Stocke1991,Maccacaro1994} to make the first
detection of evolution in the observed XLF.  Based on a steepening of
the high-redshift luminosity function, they found a deficit of
high-luminosity clusters at $z>0.3$ with a statistical significance of
approximately 3$\sigma$.  Though most subsequent investigations
corroborate these findings, some questions have been raised
concerning the reliability of the EMSS evolution detection
\citep[e.g.,][]{Nichol1997,Ebeling2000a,Ellis2002,Lewis2002}.  Of
historical note, the only other pre-{\em ROSAT\/} measurement of
significant cluster evolution came from \citet{Edge1990} who found
rapid negative evolution in the luminosity function at $z<0.2$ based
on a {\em HEAO-1} sample.  This was ultimately overruled by a
definitive and non-evolving measure of the local XLF
\citep{Ebeling1997}.

Seeking in part to confirm or refute the EMSS's controversial claim of
negative cluster evolution, a large number of cluster surveys were
initiated in the 1990s based on {\em ROSAT\/} data
\citep{Voges1992,Truemper1993}.  The 160 Square Degree {\em ROSAT\/}
Cluster Survey \citep[hereafter 160SD,][]{Vikhlinin1998,Mullis2003a}
is one such program and the subject of this paper.  Additional surveys
probing to high redshift include the NEP
\citep{Mullis2001b,Henry2001,Gioia2003}, BMW-HRI
\citep{Moretti2001, Panzera2003}, BSHARC \citep{Romer2000}, MACS
\citep*{Ebeling2001c}, RDCS \citep{Rosati1995,Rosati1998,Rosati2000}, RIXOS
\citep{Castander1995, Mason2000}, SSHARC \citep{Burke2003}, and WARPS
\citep{Scharf1997,Perlman2002}.  Complementary surveys at low
redshifts ($z \la 0.3$) yielded accurate determinations of the local
luminosity function, thus providing the crucial baseline for the
detection of redshift evolution.  These local surveys include the
BCS+eBCS \citep{Ebeling1998,Ebeling2000b}, RASS1BS
\citep{DeGrandi1999b}, and REFLEX \citep{Boehringer2001}. From the
shear number of projects it should be clear that {\em ROSAT\/} was a
watershed event for X-ray cluster surveys.

Critical comparisons of the {\em ROSAT\/} and EMSS luminosity
functions must account for the overlap (or lack thereof) of the
measurements in redshift and luminosity.  Insufficient attention to
this point led to confusion in some early analyses and debates.
Different combinations of sensitivity and areal coverage in
flux-limited surveys, when convolved with the intrinsic cluster
luminosity function, result in populating different regions of the
observed luminosity-redshift plane (see indicative results in
\mbox{Figure \ref{fig:lxzall}}).  In practice it is difficult to
directly test the cluster evolution seen in the EMSS detection
because it lies at the extreme bright end of the luminosity function.
Thus large search volumes are required to detect adequate numbers of
such rare clusters.  The 160SD is one of the largest serendipitous X-ray
surveys conducted with {\em ROSAT}.  With this substantial areal
coverage and high sensitivity, the 160SD survey is
well-positioned to probe cluster evolution.  Preliminary analyses of
our sample, with at-the-time incomplete optical follow-up, seemed to
confirm the deficit of high-luminosity, high-redshift clusters first
seen by the EMSS \citep{Vikhlinin1998b,Vikhlinin2000}.

%................................................................
% FIGURE 2 --- IDL: /figs/fig2-skycov.pro ---
\begin{figure}
\epsscale{1.1}
\plotone{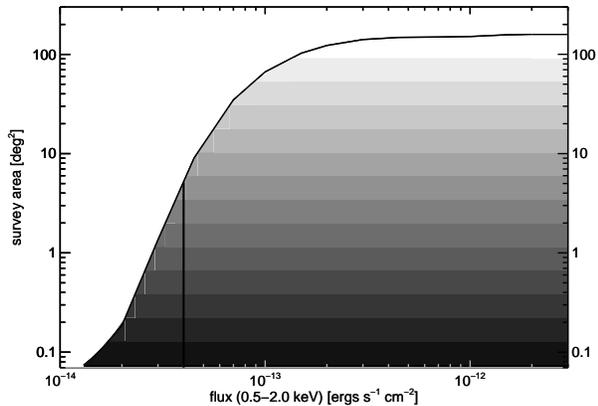}
\caption{The selection function of the 160SD survey specifying the
square degrees of area covered as a function of limiting X-ray flux
\mbox{(0.5--2.0\,keV)}.  These data are tabulated in \citet[][Table
5]{Vikhlinin1998}.  The shading encodes this information for subsequent
use in \mbox{Figure \ref{fig:lxz}}.  The vertical line indicates the
minimum flux limit (\mbox{4 $\times$ 10$^{-14}$} \fxunits,
\mbox{0.5--2.0\,keV}) used in the present
analysis. \label{fig:skycoverage}}
\end{figure}
%................................................................

In this paper we present measurements of the cluster XLF out to
$z=0.8$ and describe their implications for cluster evolution based on
the final 160SD sample of 201 clusters.  In \S\,2 we outline the basic
construction of our cluster sample and the associated selection
function used herein.  The formalism associated with the XLF is
defined in \S\,3 and used to measure cluster abundances at both low
and high redshifts.  In \S\,4 we characterize the evolution in the
cluster population using integrated number counts and a
maximum-likelihood analysis of the observed luminosity-redshift
distribution relative to an evolving Schechter function.  We discuss
our results in the context of previous works and draw conclusions in
\S\,5.  Throughout this analysis we use the cosmological parameters
$H_{0} = 50$ \hpone km s$^{-1}$ Mpc$^{-1}$, $\Omega_{M}=1$, and
$\Omega_{\Lambda} = 0$ (Einstein--de-Sitter, EdS) for direct
comparison to previous work in this field.  We also repeat
calculations in the currently preferred cosmology where
$\Omega_{M}=0.3$ and $\Omega_{\Lambda} = 0.7$. \mbox{X-ray} fluxes and
restframe luminosities are quoted in the {\mbox 0.5--2.0 keV} energy
band unless otherwise stated.

%--------------- SECTION 2: THE CLUSTER SAMPLE ------------------
\section{The 160SD Cluster Sample}
\label{The 160SD Cluster Sample}

The 160SD sample of 201 galaxy clusters is the largest high-redshift,
X-ray selected sample published so far.  For instance, there are 73
objects at $z>0.3$ and 22 objects at $z>0.5$.  The 160SD sample was
constructed via the serendipitous detection of extended X-ray sources
in 647 archival {\em ROSAT\/} PSPC observations.  Of 223 cluster
candidates, we identified 201 as galaxy clusters, 21 as probable
false-detections due to blends of unresolved point sources, and one
source is unidentified due to its proximity to a bright star.  We have
secured spectroscopic redshifts for 200 of the 201 clusters.
\citet{Vikhlinin1998} give a complete description of the survey
methodology.  \citet{Mullis2003a} detail the optical follow-up and
present the final cluster catalog with spectroscopic redshifts.  Note
that the number of false-detections in the sample agrees very well
with that expected from the confusion of point sources as demonstrated
in the Monte-Carlo simulations of \citet{Vikhlinin1998}.  Two
false-detections (Nos.\ 77 and 141) serendipitously imaged by {\em
XMM-Newton\/} and {\em Chandra}, respectively, are indeed confused
point sources.  Finally, the optical survey imaging was sufficiently
deep to demonstrate that none of the false-detections should be galaxy
clusters at $z \la 0.9$.

\begin{figure*}
\epsscale{1}
\plotone{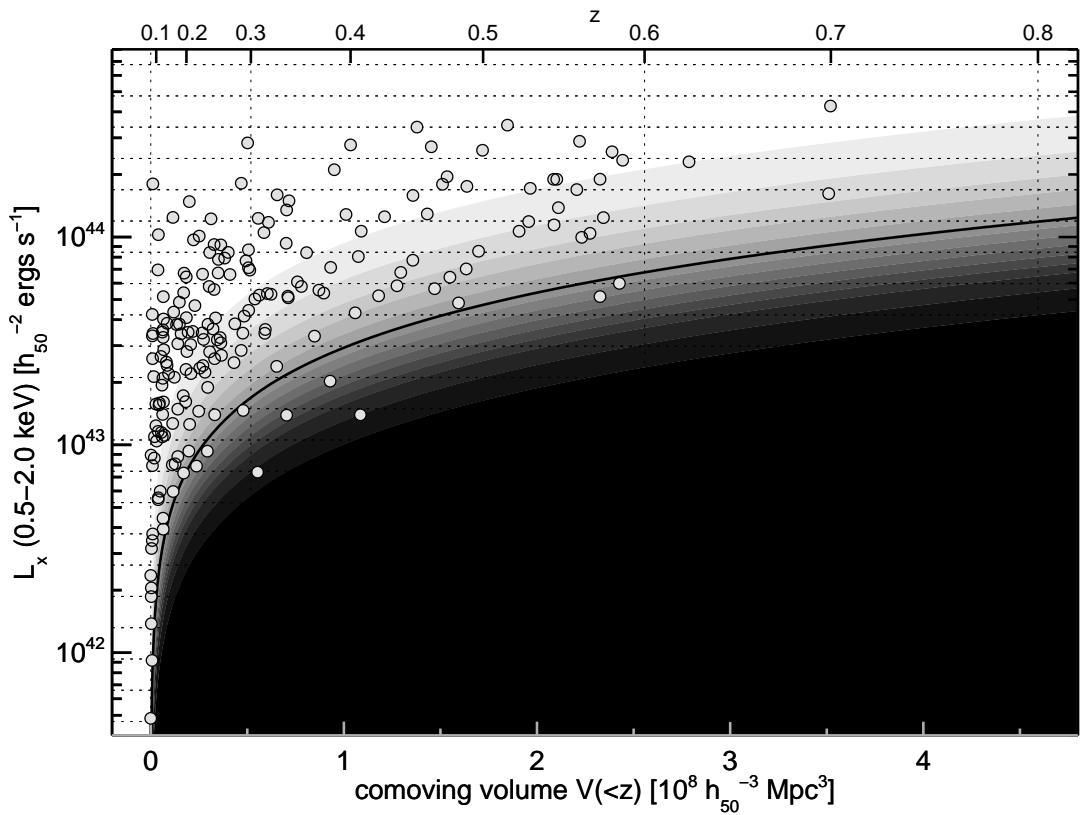}
\caption{X-ray luminosity of the 160SD clusters versus redshift $z$
({\em top axis}) and comoving volume out to redshift $z$ ({\em bottom
axis}).  The volume corresponds to a survey solid angle of 158.5
deg$^{2}$ in an Einstein--de-Sitter cosmology.  The shading,
established in \mbox{Figure \ref{fig:skycoverage}}, demonstrates the range of
survey flux limits and associated sky coverage in this parameter space.
The thick, solid curve indicates the minimum flux limit (\mbox{4
$\times$ 10$^{-14}$} \fxunits, \mbox{0.5--2.0\,keV}) used in the
present analysis.  The horizontal (vertical) dashed lines are the
boundaries of the luminosity (redshift) bins used in computing the
luminosity function.  
%Underlining the observed distribution are
%logarithmically-spaced isocontours of the integrated number of
%clusters expected to be observed assuming a non-evolving XLF.
\label{fig:lxz}}
\end{figure*}

Although spatial X-ray extent was our primary selection criterion,
detailed comparisons with other surveys demonstrate that no known
clusters were missed as unresolved sources \citep[see \S\,4 of][and
references therein]{Mullis2003a}.  Thus the 160SD clusters are in
effect drawn from a statistically complete, flux-limited survey with
an areal coverage ($\Omega$) and sensitivity characterized by the
selection function shown in Figure \ref{fig:skycoverage}.  A total of
158.5 deg$^{2}$ were surveyed and the median survey flux (where
$\Omega = \Omega _{\rm total} / 2$) is \mbox{1.2 $\times$ 10$^{-13}$}
\fxunits \,\mbox{(0.5--2.0\,keV)}.  We restrict our analysis to a
minimum flux of ~\mbox{4 $\times$ 10$^{-14}$} \fxunits~ to avoid any
uncertainties in the sky coverage at very faint fluxes.  There are 190
clusters in the 160SD survey sample above this flux limit.
\mbox{Figure \ref{fig:lxzall}} shows the position in
luminosity-redshift space of our sample relative to the BCS, one of
the key reference samples at low redshift, and the EMSS.  \mbox{Figure
\ref{fig:lxz}} shows the 160SD data in greater detail. Here the
cluster luminosities are plotted as a linear function of comoving
volume which provides a more uniform view of the volume sampling.  The
greyscale in the figure indicates the parameter space probed by our
survey --- sensitivity is maximal in light regions and minimal in dark
regions.

In our subsequent derivations of the cluster XLF and tests for
evolution, we exclude several objects (all at $z<0.3$) to minimize the
biasing of our results.  To avoid potentially skewing the impartiality
of the sampling, we reject nine clusters whose redshifts are within
$\Delta z = 0.015$ of the original target of the {\em ROSAT PSPC\/}
observations (Nos.\ 16, 32, 112, 134, 165, 166, 174, 177, 206). Four
X-ray-overluminous elliptical galaxies or ``fossil groups'' (Nos.\
110, 144, 201, and 211) are discounted because these special
structures are unlikely to meet the selection criteria of the local
samples \citep{Vikhlinin1999b}. Thus the final cluster sample used
here consists of 177 clusters.

%.............................................................
% FIGURE 3 --- IDL: /figs/fig3-lxz.pro ---
%................................................................

%--------------- SECTION 3: THE XLF --------------------------------------
\section{The X-ray Luminosity Function}
\label{The X-ray Luminosity Function}

We define the cluster differential luminosity function to be 

\begin{equation}
\phi(L_{\rm X}, z)~= \frac{d^{2}N}{dVdL_{\rm X}}(L_{\rm X},z) 
\label{basicxlf}
\end{equation}

\noindent where $N$ is the number of clusters of luminosity $L_{\rm
X}$ in a volume $V$ at a redshift $z$.  The standard approach for
deriving a nonparametric representation of the differential cluster
XLF is based on the $1/V_{\rm max}$ technique first proposed by
\citet{Schmidt1968} and generalized by \citet{Avni1980}. Here the
observed luminosity range is parsed into bins of width $\Delta L$,
within each of which there are $N_{j}$ observed clusters.  The XLF is
estimated by summing the density contributions of each cluster in the
considered luminosity/redshift bin,

\begin{equation}
\phi(L_{{\rm X}},z)~=
   ~\frac{1}{\Delta L} \sum_{i=1}^{N_j}~\frac{1}{V_{\rm max}(L_{{\rm X}, i})~} ,
\label{xlf}
\end{equation}

\noindent where $V_{\rm max}$ is the total comoving volume in which a
cluster of luminosity $L_{{\rm X}, i}$ could have been detected above
the flux limits of the survey.  Over a specific redshift interval
($z_{\rm min} < z < z_{\rm max}$), this search volume is defined to
be

\begin{equation}
V_{\rm max}(L_{\rm X})~=~\int_{z_{\rm min}}^{z_{\rm max}} \Omega(f_{\rm
X}(L_{\rm X},z))~\frac{dV(z)}{dz}~dz.
\label{va}
\end{equation}

\noindent Here $\Omega(f_{\rm X})$ is the sky area surveyed in
 steradians as a function of X-ray flux and $\frac{dV(z)}{dz}$ is the
 differential, comoving volume element per steradian.   

\citet{Page2000} describe a refinement to the canonical approach in which
the estimator takes the form

\begin{equation}
\phi(L_{{\rm X}},z)~=
   ~\frac{N_{j}}{\int_{L_{\rm X, min}}^{L_{\rm X, max}}
                 \int_{z_{\rm min}}^{z_{\rm max}}
                 \Omega(f_{\rm X}(L_{\rm X},z))~\frac{dV(z)}{dz}~dz~dL_{\rm X}} 
\label{pagexlf}
\end{equation}
 
\noindent where the boundaries of the luminosity bin are specified by
$L_{\rm X, min}$ and $L_{\rm X, max}$.  The effect of bringing the
luminosity interval into the double integral results in a better
estimate of the {\em effective} $\Delta L$ which can be smaller than
the full bin width for regions of luminosity-redshift space transected
by the survey flux limit (e.g., faint luminosities).  We will use the
Page-Carrera (PC) estimator in our subsequent derivations of the
cluster XLF.  However, the results are essentially identical to the
classical $1/V{\rm max}$ procedure except at the very faint end where
the XLF is marginally increased.

As for a parametric representation of the cluster XLF, observations
are well fit by a Schechter function \citep{Schechter1976} of
the form

\begin{equation}
\phi(L_{{\rm X}},z)\, dL_{\rm X}~=~\phi^{\star} 
   \left( \frac{L_{\rm X}}{L_{\rm X}^{\star}} \right)^{- \alpha}
   \exp \left(- \frac{L_{\rm X}}{L_{\rm X}^{\star}} \right) 
   \frac{dL_{\rm X}}{L_{\rm X}^{\star}}
\label{eq:Schechter}
\end{equation}

\noindent where $\phi^{\star}$ is the normalization (units $h_{50}^3$
Mpc$^{-3}$), $\alpha$ is the faint-end slope, and $L_{\rm X}^{\star}$
is the characteristic luminosity marking the interface between the
power-law and the exponential regimes.
An equivalent expression for the XLF commonly used in the literature
is

\begin{equation}
\phi(L_{{\rm X}},z)\,~=~{\cal A}\, L_{\rm X}^{- \alpha}
   \exp \left(- \frac{L_{\rm X}}{L_{\rm X}^{\star}} \right) .
\end{equation}

\noindent With $L_{\rm X}$ in units of \mbox{10$^{44}$ $h_{50}^{-2}$}
\lxunits, the associated normalization, ${\cal A}$, has units of
\mbox{$h_{50}^3$ Mpc$^{-3}$} \mbox{(10$^{44}$ $h_{50}^{-2}$
\lxunits)$^{\alpha -1}$}, and the two normalizations are related by
$\phi^{\star}\,=\,{\cal A}\,(L_{\rm X}^{\star})^{1-\alpha}$.

%...................................................................
% TABLE 1 --- XLF Schechter Fits
\input{tab1.tex}

\newpage

\subsection{The Local XLF}

Knowledge of the local or near present-day abundance of clusters is
fundamental to evolutionary studies because it serves as the
no-evolution baseline against which distant cluster samples can be
tested.  As previously noted, one of the significant achievements of
the {\em ROSAT\/} era is the accurate determination of the local
cluster XLF ($z \la 0.3$).  Three principal measurements are based on
the BCS, RASS1BS, and REFLEX samples which were constructed by
surveying large portions of the two-thirds of the sky outside the
Galactic plane\footnote{Note that the significant gap in all-sky
coverage due to the former zone of avoidance ($|b| < 20^{o}$) is being
redressed by the CIZA survey \citep*{Ebeling2002}} at relatively
bright fluxes. \citet{Ebeling1997} reported the first results based on
199 BCS clusters in the northern hemisphere ($f_{\rm X} >$ \mbox{2.8
$\times$ 10$^{-12}$} \fxunits, \mbox{0.1--2.4\,keV}). Part of a pilot
program of the larger REFLEX survey, \citet{DeGrandi1999a} presented a
measurement based on 129 RASS1BS clusters ($f_{\rm X} >$ \mbox{3--4
$\times$ 10$^{-12}$} \fxunits, \mbox{0.5--2.0\,keV}) from the south
Galactic cap.  Finally, the XLF determination based on the largest
sample to date comes from the REFLEX survey.  \citet{Boehringer2002}
made a detailed analysis of 452 clusters extracted from the southern
celestial hemisphere ($f_{\rm X} >$ \mbox{3 $\times$ 10$^{-12}$}
\fxunits, \mbox{0.1--2.4\,keV}).

\begin{figure*}
\epsscale{0.95}
\plotone{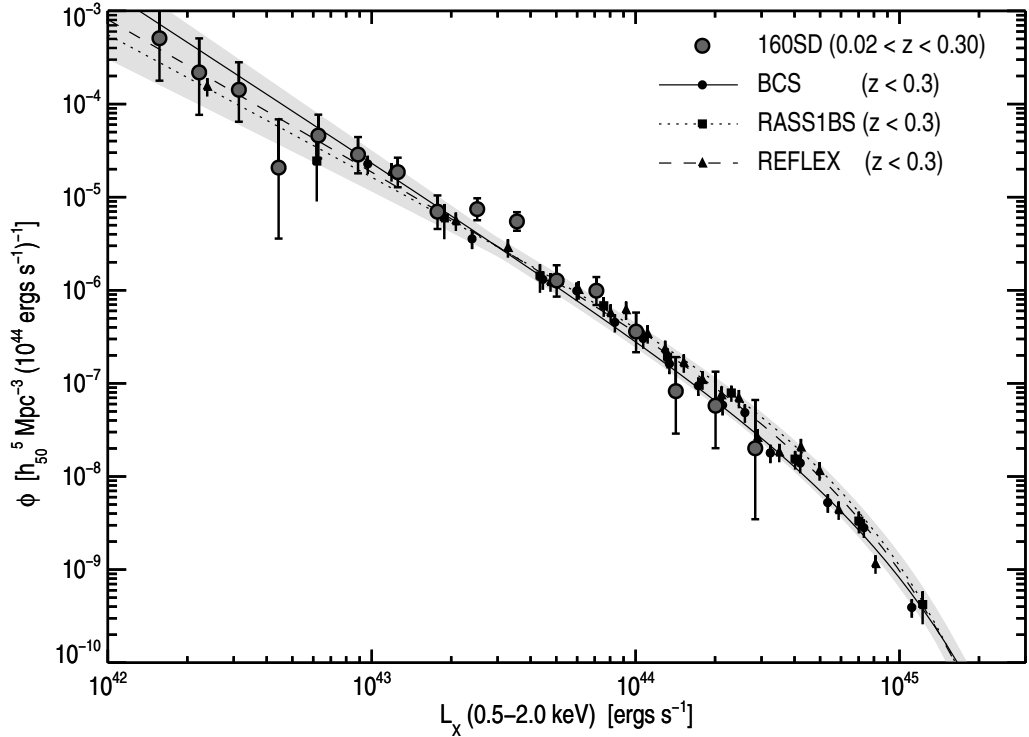}
\caption{Determinations of the local cluster X-ray luminosity function
as measured by the 160SD survey and the local reference samples (BCS,
RASS1BS, and REFLEX) in an Einstein--de-Sitter universe.  The 160SD
data values along with the number of clusters and average cluster
redshift for each luminosity bin are given in Table
\ref{tab:lozxlf}. Nonparametric data points and Schechter fits for the
reference samples are from \citet{Ebeling1997}, \citet{DeGrandi1999a},
and \citet{Boehringer2002}, respectively.  The BCS data points are
based on a merged analysis of the BCS+eBCS samples (H. Ebeling 2003,
private communication). The shaded region indicates the $1\sigma$
uncertainty envelope of the Schechter fits assuming the errors on
$L_{\rm X}^{\star}$ and $\alpha$ are correlated. The indicated
uncertainties on data points are $\pm1\sigma$.\label{fig:localxlf}}
\end{figure*}

Nonparametric determinations of the local XLF ($z < 0.3$) from the
all-sky samples (BCS, RASS1BS, and REFLEX) are shown in Figure
\ref{fig:localxlf}.  We also plot the best-fitting Schechter functions
for these data and list the associated best-fit parameters in
\mbox{Table \ref{tab:schechter}} for future reference.  The results
demonstrate that we have accurate knowledge of the local cluster
luminosity function.  Independent investigators using different X-ray
selection procedures over different regions of the sky agree on the
local space density of clusters.  For example, in the luminosity
interval \mbox{5 $\times$ 10$^{43}$ -- 10$^{45}$} \lxunits~
\mbox{(0.5--2.0\,keV)}, the internal accuracy of these XLF
measurements is approximately $\pm$10\%--20\% (estimated from the
$\pm1\sigma$ excursion of the error envelopes plotted in Figure
\ref{fig:localxlf}).  Moreover, the systematics are also small --- the
results from the BCS and RASS1BS surveys vary a maximum of about
$\pm$25\% relative to the Schechter fit of REFLEX.

\begin{figure*}
\epsscale{0.95}
\plotone{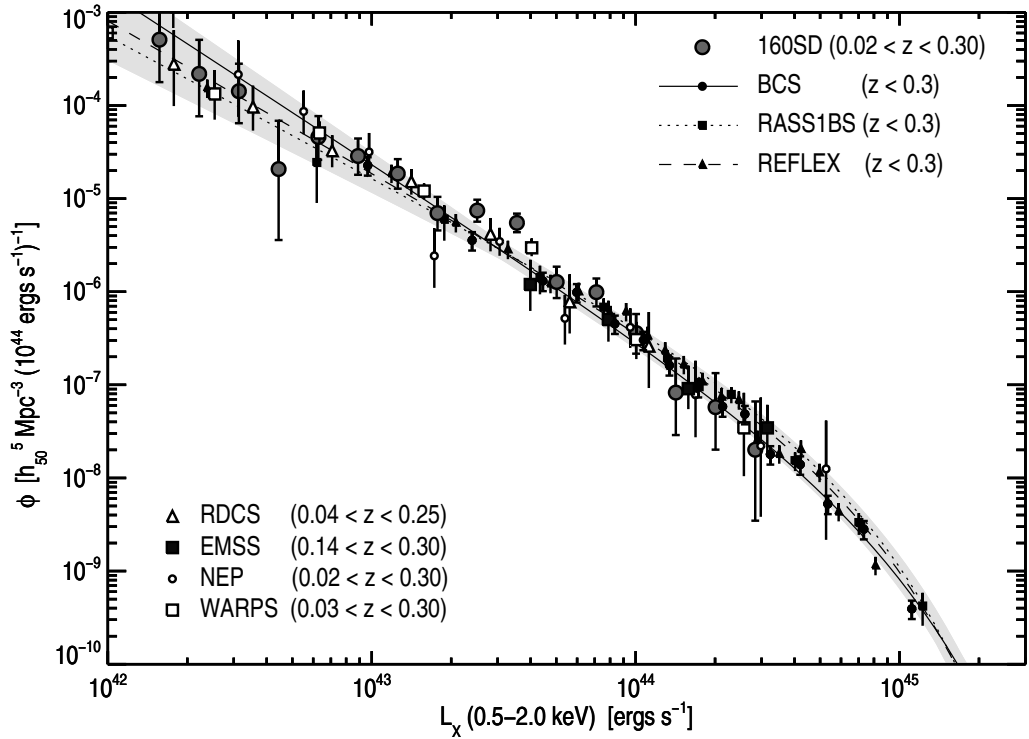}
\caption{Compilation of local XLFs as measured by eight X-ray
flux-limited surveys.  RDCS: \citet{Rosati1998}, EMSS:
\citet{Henry1992}, NEP: \citet{Gioia2001}, and WARPS:
\citet{Jones2000c} and the references in Figure 4 (Einstein--de-Sitter
universe).\label{fig:localxlfb}}
\end{figure*}

Before considering the cluster population at high redshift, we first
examine the low-redshift diagnostic using the 160SD cluster sample. 
%check the reliability of our 160SD cluster sample by deriving the
%low-redshift diagnostic.  
Using the procedure previously described, we estimate the local XLF
between \mbox{$\sim$10$^{42}$ and 3 $\times$ 10$^{44}$} \lxunits~
using the 110 clusters at $0.02 < z < 0.3$ in the 160SD survey.  Our
measurement is plotted in Figure \ref{fig:localxlf} and is tabulated
in Table \ref{tab:lozxlf}. The luminosity binning is uniform in log
space and data points are plotted at the center of the luminosity
interval.  The error bars are the equivalent $\pm1\sigma$
uncertainties based on Poissonian errors for the number of clusters
per bin \citep{Gehrels1986}.  Note the good agreement at low redshift
between the 160SD and the three principal local samples.  In Figure
\ref{fig:localxlfb} we add the measurements from four additional deep
surveys (RDCS, EMSS, NEP, and WARPS) thus producing a compilation of
all of the {\em ROSAT\/} (plus EMSS) local XLFs published to date from
X-ray selected, X-ray flux-limited cluster samples.  The shallow
all-sky surveys accurately measure the local luminosity function at
intermediate to high luminosities ($\ga 10^{43}$ \lxunits) but are
relatively insensitive to very low luminosity clusters ($\la 10^{43}$
\lxunits).  In a complementary fashion, the deep surveys better
measure the faint end of the local XLF, provide reasonable precision
at intermediate luminosities, but poorly constrain the very bright end
($\ga 10^{44}$ \lxunits) due to limited survey volumes at low
redshift.  The cluster luminosity-redshift distributions in
\mbox{Figure \ref{fig:lxzall}} also illustrate this dependence on flux
limit and solid angle.

%...................................................................

\subsection{The High-Redshift XLF}

\begin{figure*}
\epsscale{.95}
\plotone{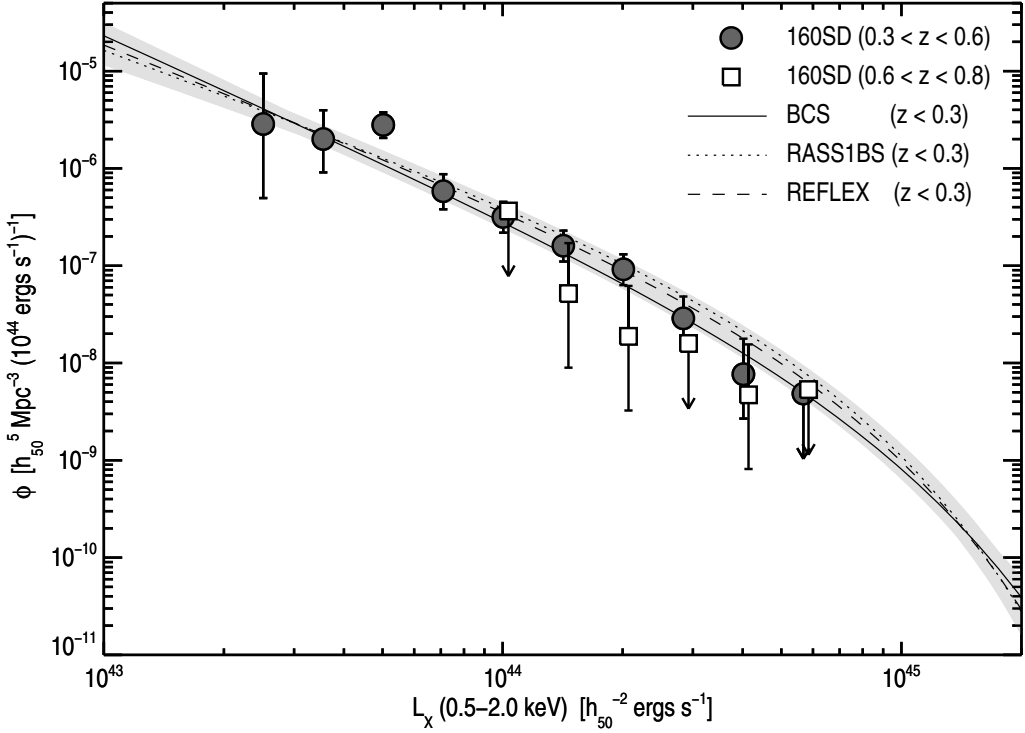}
\caption{The high-redshift cluster X-ray luminosity function from the
160SD sample (Einstein--de-Sitter universe).  The data points for $0.6
< z < 0.8$ have been slightly offset to avoid confusion.  The 160SD
data values along with the number of clusters and average cluster
redshift for each luminosity bin are given in \mbox{Table
\ref{tab:xlf}}.  Schechter fits to the
local XLF are also plotted as described in Figure
\ref{fig:localxlf}. If no clusters are detected in a luminosity bin, we plot the $1\sigma$ upper limit for the Poisson
error on zero clusters. \label{fig:hizxlf}}
\end{figure*}

We measure the distant XLF using the 66 clusters from the 160SD sample
at $0.3 < z < 0.8$ and fluxes above \mbox{4 $\times$ 10$^{-14}$}
\fxunits.  The numerical results are shown in \mbox{Table
\ref{tab:xlf}} and plotted in
\mbox{Figure \ref{fig:hizxlf}}.
We have probed a sufficiently large volume such that we can derive
useful results in two intervals: \mbox{$0.3 < z < 0.6$} and \mbox{$0.6
< z < 0.8$}.  
In addition to our high-redshift measurements, we also show in Figure
\ref{fig:hizxlf} the local determinations of the XLF.  These establish
the regime where the high-redshift results should lie if the spatial
density of clusters does not evolve out to the considered redshifts.

%................................................................
% FIGURE 4 --- IDL: /figs/fig45-localxlf.pro ---
%................................................................

%................................................................
% FIGURE 5 --- IDL: /figs/fig45-localxlf.pro ---
%................................................................

%................................................................
% FIGURE 6 --- IDL: /figs/fig67-hizxlf.pro ---
%................................................................

%................................................................
% FIGURE 7 --- IDL: /figs/fig67-hizxlf.pro ---
\begin{figure*}
\epsscale{.95}
\plotone{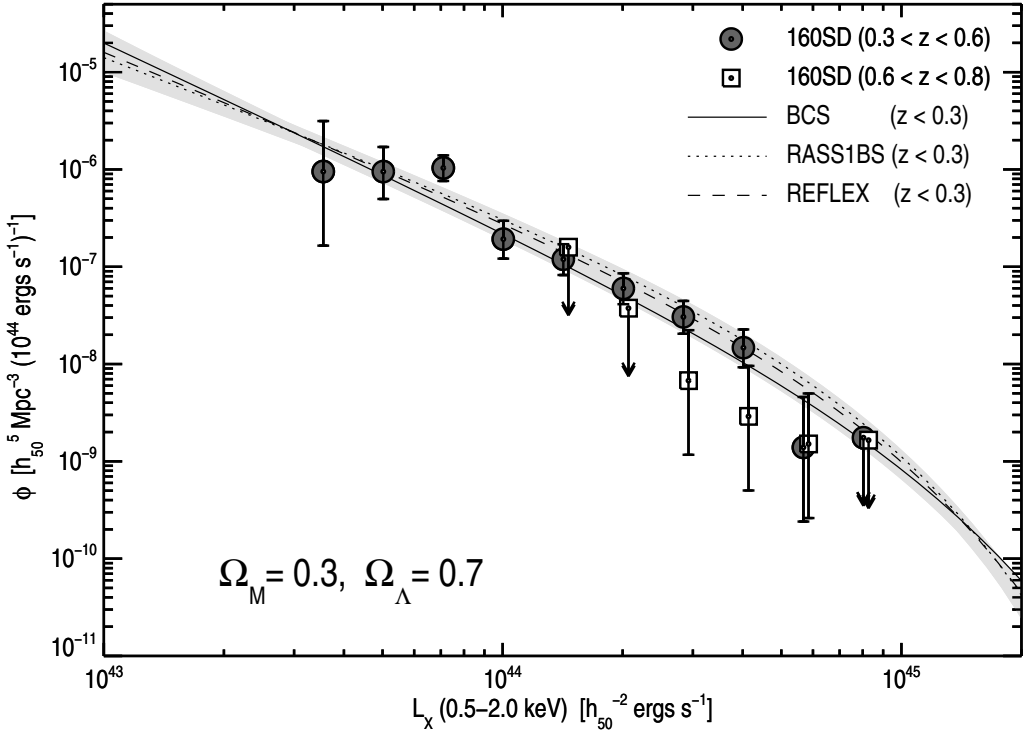}
\caption{The high-redshift cluster X-ray luminosity function from the
160SD sample for a cosmology characterized by the parameters
$\Omega_{M}=0.3$ and $\Omega_{\Lambda} = 0.7$.  The data points for
$0.6 < z < 0.8$ have been slightly offset to avoid confusion. The
160SD data values along with the number of clusters and average
cluster redshift for each luminosity bin are given in \mbox{Table
\ref{tab:xlf_lcdm}}.  The REFLEX
XLF was explicitly measured for a $\Lambda$-dominated cosmology by
\citet{Boehringer2002}.  We transform the BCS and RASS1BS results to
this cosmology based on the REFLEX results (see text for
details). \label{fig:hizxlf_lcdm}}
\end{figure*}
%................................................................

% TABLE 2 -- 160SD low-z (Eds)
% ---- tab2.tex      = HEADER section
% ---- tab2-data.tex = IDL: xlf2latex.pro via call from fig45-localxlf.pro ---
\input{tab2.tex}

Our measurement of the XLF at \mbox{$0.3 < z < 0.6$} probes the
luminosity range \mbox{2 $\times$ 10$^{43}$ -- 7 $\times$ 10$^{44}$}
\lxunits~ (\mbox{Figure \ref{fig:hizxlf}}: filled circles). Except for
the data point near \mbox{5 $\times$ 10$^{43}$ \lxunits} which is
$1.9\sigma$ off the local relation, there is excellent agreement
between these results and the no-evolution benchmark at least out to
intermediate luminosities, \mbox{$L_{\rm X} \sim$ 2 $\times$
10$^{44}$} \lxunits.  In this region the 160SD best matches the
normalization of the REFLEX XLF.  At higher luminosities the distant
XLF is lower than the local population but we will demonstrate in
\S\,\ref{Integrated Number Counts} that this effect is only marginally
significant with respect to the local XLF with the highest
normalization (RASS1BS).  The median redshift for these depressed data
points is $<z> = 0.50$ (see \mbox{Table \ref{tab:xlf}}).

At the highest redshifts probed by the 160SD in the present analysis,
\mbox{$0.6 < z < 0.8$}, we measure useful constraints over the
luminosity interval \mbox{1~--~6 $\times$ 10$^{44}$} \lxunits, and the
median cluster redshift is $<z> = 0.70$.  (\mbox{Figure
\ref{fig:hizxlf}}: open squares).  The distant cluster volume
densities are below the local level at all measured luminosities.  We
will show that this result is significant ($>3\sigma$) in both the
Einstein--de-Sitter and $\Lambda$-dominated models even with respect
to the lowest-normalization local XLF (\S\,\ref{Integrated Number
Counts}).  This is a deficit of high luminosity, high redshift
clusters in a manner similar to that seen in the EMSS results.  Note
that although the optical CCD imaging used to identify the 160SD
clusters was sufficiently deep to reveal massive clusters to $z=0.9$,
we have conservatively cut off the most distant redshift shell at
$z=0.8$.  This lessens any potential negative evolution in the results
since the larger redshift boundary would increase the search volume
without adding clusters and hence depress the data points.

To assess the impact of changing the cosmological framework from an
Einstein--de-Sitter to a $\Lambda$-dominated universe, we repeat our
calculation of the cluster luminosity function setting
$\Omega_{M}=0.3$ and $\Omega_{\Lambda} = 0.7$, the results of which
are shown in \mbox{Figure \ref{fig:hizxlf_lcdm}} and \mbox{Table
\ref{tab:xlf_lcdm}}.  The Einstein--de-Sitter XLF ($\phi_{\rm EdS}$,
\mbox{Figure \ref{fig:hizxlf}}) and $\Lambda$-dominated XLF
($\phi_{\Lambda}$) appear very similar because the data points of the
latter are offset diagonally down and to the right approximately along
the former.  This is the combined effect of the increase in both the
cluster luminosities and the search volumes in a $\Lambda$-dominated
universe.
The actual positioning
of the two XLFs relative to each other depends on the cluster
redshifts and the specific luminosity interval.  For example in the
REFLEX survey ($<z>=0.08$), between 10$^{43}$ and 10$^{45}$ \lxunits~
the ratio of the fitted XLFs ($\phi_{\Lambda}$/$\phi_{\rm EdS}$) is
less than unity with a broad minimum of $\sim$0.8 around
\mbox{10$^{44}$} \lxunits.  REFLEX is the only local sample for
which the XLF in a $\Lambda$-cosmology has been explicitly measured
\citep{Boehringer2002}.  However, given the similarity of the redshift
and luminosity distributions, we have used the ratio
$\phi_{\Lambda}$/$\phi_{\rm EdS}$ from REFLEX to make an approximate
transform of the BCS and RASS1BS to this alternate cosmology.

It is apparent from \mbox{Figure \ref{fig:hizxlf_lcdm}} that the
high-redshift 160SD XLFs and the local XLFs shift in similar ways;
thus the apparent deficit of high-luminosity clusters persists in the
$\Lambda$-dominated cosmology.  The only important difference, of
course, is that the point where the 160SD data depart significantly
from the non-evolution baseline is shifted to larger luminosities, and
this is anticipated given the increase in luminosity distance
\mbox{($L_{\rm X} \ga$ 3 $\times$ 10$^{44}$} \lxunits~ for
\mbox{\mbox{$0.3 < z < 0.6$}}, and \mbox{$L_{\rm X} \ga$ \mbox{2
$\times$ 10$^{44}$}} \lxunits~ for \mbox{$0.6 < z < 0.8$}).

We will examine the significance and strength of this apparent cluster
evolution in the following section.

% TABLE 3 -- 160SD hiz-z (Eds)
% ---- tab3.tex      = HEADER section
% ---- tab3a-data.tex = IDL: xlf2latex.pro via call from fig67-hixlf.pro ---
% ---- tab3b-data.tex = IDL: xlf2latex.pro via call from fig67-hixlf.pro ---
\input{tab3.tex}
%\mbox{}\pagebreak

% TABLE 4 -- 160SD hiz-z (LCDM)
% ---- tab4.tex      = HEADER section
% ---- tab4a-data.tex = IDL: xlf2latex.pro via call from fig67-hixlf.pro ---
% ---- tab4a-data.tex = IDL: xlf2latex.pro via call from fig67-hixlf.pro ---
\input{tab4.tex}

%--------------- SECTION 4: QUANTIFYING EVOLUTION ------------------
\section{Quantifying Evolution}

X-ray luminosity functions like those shown in \mbox{Figures
\ref{fig:localxlf}--\ref{fig:hizxlf_lcdm}} are useful for
visualizations and qualitative assessments of the cluster population;
however, they are non-optimal for quantitative analyses.  For example
ambiguities exist in the selection of the luminosity binning (e.g.,
fixed or adaptive intervals) and the loci of the plotted data points
in luminosity space (e.g., at the bin center or the density-weighted
mean luminosity).  Furthermore, for the case of negative evolution the
effect that we are attempting to measure is either a diminishing
signal or a non-detection.  Thus we will apply the alternate approaches
of integrated number counts and a maximum-likelihood fit of an
evolving model XLF to quantify the significance and strength of the
apparent evolution in the 160SD clusters.

% TABLE 5 -- Integrated Number Counts  
% ---- tab5.tex      = HEADER section
% ---- tab5a-data.tex = IDL: misc.pro --> d160sigtable.pro
% ---- tab5b-data.tex = IDL: misc.pro --> d160sigtable.pro
% ---- tab5c-data.tex = IDL: misc.pro --> d160sigtable.pro
% ---- tab5d-data.tex = IDL: misc.pro --> d160sigtable.pro
\input{tab5.tex}

%--------------- SECTION 4a: INTEGRATED NUMBER COUNTS ------------------
\newpage
\subsection{Integrated Number Counts}
\label{Integrated Number Counts}

For a given region of luminosity-redshift space we compare the number
of observed clusters ($N_{\rm obs}$) with the number that are expected
($N_{\rm exp}$) assuming there is no evolution in the population.  The
latter is computed by integrating the local luminosity function,
$\phi(L_{\rm X},z$), over luminosity and redshift, and folding this
through the survey selection function, $\Omega(f_{\rm X}$), using the
following equation,

\begin{equation}
N_{\rm exp}~=~\int_{L_{\rm X, min}}^{L_{\rm X, max}}
                 \int_{z_{\rm min}}^{z_{\rm max}}
                 \phi(L_{\rm X},z)~\Omega(f_{\rm X}(L_{\rm X},z))~\frac{dV(z)}{dz}~dz~dL_{\rm X}.
\label{eq:counts}
\end{equation}

\noindent Note that the {\em non-evolving} XLF is strictly a function
of luminosity as parameterized by the Schechter fits to local clusters
(e.g., \mbox{Equation \ref{eq:Schechter}} with the best-fit parameters
from \mbox{Table \ref{tab:schechter}}).  However, we explicitly
indicate the potential redshift dependence in this equation to
generalize it for subsequent treatment with an evolving XLF.  The
statistical significance of the difference between $N_{\rm obs}$ and
$N_{\rm exp}$ is computed based on Poisson confidence intervals
\citep{Gehrels1986}.

Of the three local XLFs (BCS, RASS1BS, and REFLEX), we use the REFLEX
measure as the preferred reference for three reasons: 1) the REFLEX
local XLF is based on the largest sample of clusters used to date (452
clusters), 2) the REFLEX normalization lies intermediate to the BCS
and RASS1BS, and 3) at low to intermediate luminosities, our 160SD
low-redshift data most closely match the REFLEX normalization.
Nonetheless, we will also quote our results relative to the BCS and
RASS1BS to demonstrate the full range of possible significances.

First we consider the redshift interval \mbox{$0.3 < z < 0.6$}
starting with the highest luminosity bin and then integrating to lower
luminosities in an Einstein--de-Sitter model. In and above the highest
bin of the XLF ($L_{\rm X} \ga 4.8$), we observe zero clusters whereas
the REFLEX local XLF predicts 4.5 clusters according to \mbox{Equation
\ref{eq:counts}}.  This difference is only 2.3$\sigma$ significant.
The next two bins ($L_{\rm X} \ga 3.4, 2.4$) have $N_{\rm obs}=2,7$
and $N_{\rm exp}=9.1,16.2$ with the associated significances of
2.6$\sigma$ and 2.4$\sigma$.  Continuing from here to lower luminosity
bins decreases the significance.  These results are summarized in
\mbox{Table \ref{tab:counts}}.  Examining the run of significance
versus minimum luminosity without the constraints of the arbitrary
bins of \mbox{Figure \ref{fig:hizxlf}} indicates that the significance
briefly peaks above 3$\sigma$ near \mbox{3.4 $\times$ 10$^{44}$}
\lxunits.  However with the BCS XLF as the no-evolution reference, the
apparent cluster deficit is not significant (i.e. always $<3\sigma$).
Conversely, the RASS1BS XLF indicates a statistically strong signal
across much of the probed luminosity.  Repeating this analysis in the
context of a $\Lambda$-dominated universe uniformly reduces the
significance of the cluster deficit such that only relative to the
RASS1BS does the deviation seen in the 160SD measure at \mbox{$0.3 < z
< 0.6$} appear marginally significant (see \mbox{Table
\ref{tab:counts}}).

Our highest redshift measure of the XLF, \mbox{$0.6 < z < 0.8$}, is
clearly lower than all the determinations of the local XLF
(\mbox{Figures \ref{fig:hizxlf} \& \ref{fig:hizxlf_lcdm}}).
Integrating over the entire luminosity range sampled in this redshift
shell, a non-evolving model of the cluster population based on the
REFLEX XLF predicts 22.1 clusters.  This strongly conflicts with the
actual observed sample of 3 clusters, a 4.9$\sigma$ difference.  The
situation in the $\Lambda$-cosmology is essentially the same; 21.8
clusters expected, a 4.8$\sigma$ difference.  If we instead use the
RASS1BS XLF as our baseline, the predicted count is greater than 26
clusters with the resulting cluster deficit being 5.5$\sigma$.  More
importantly, if we adopt the most conservative position (i.e.,
attempting to minimize the deficit) and use the BCS XLF, we expect to
find 16 clusters which is 3.8$\sigma$ away from the observed
value.  An integrated bin-by-bin analysis is reported in \mbox{Table
\ref{tab:counts}}.

%--------------- SECTION 4b: EVOLVING SCHECHTER FUNCTION ------------------
%\newpage
\subsection{Maximum Likelihood Analysis}
\label{Evolving Schechter Function}

A more general approach to quantifying evolution in the cluster XLF is
to perform a maximum likelihood fit of an evolving Schechter function
to the observed cluster distribution in luminosity and redshift.  This
approach makes maximal use of the data, is free from arbitrary
binning, and is sensitive to potential negative evolution
\citep{Rosati2000,Henry2002b}.

We follow the prescription of \citet{Marshall1983} but generalize the
treatment to account for uncertainties in the observations.  The
luminosity-redshift plane is uniformly parsed into extremely small
intervals of size $dL_{\rm X}dz$. In each element we
compute the expected number of clusters with luminosity $L_{\rm
X}$ and redshift $z$:

\begin{equation}
\lambda(L_{\rm X},z)dL_{\rm X}dz~=~\phi(L_{\rm X},z)~\Omega(f_{\rm X}(L_{\rm X},z))~\frac{dV(z)}{dz}dL_{\rm X}dz.
\label{eq:lambda}
\end{equation}

\noindent The model XLF, $\phi(L_{\rm X},z)$, is an evolving Schechter
function of the form given in \mbox{Equation \ref{eq:Schechter}}.  The
important modification here is to free the density normalization,
$\phi^{\star}$, and characteristic luminosity, $L_{\rm X}^{\star}$, of
the luminosity function to evolve with redshift such that,

\begin{equation}
\phi^{\star}(z)~=\phi^{\star}_{0}~\left(\frac{1+z}{1+z_0}\right)^A,
\label{eq:A}
\end{equation}

\noindent and

\begin{equation}
L_{\rm X}^{\star}(z)~=L_{{\rm X},0}^{\star}\left(\frac{1+z}{1+z_0}\right)^{B},
\label{eq:B}
\end{equation}

\noindent where $A$ and $B$ parameterize the evolution.  The local
values of the normalization ($\phi^{\star}_{0}$) and characteristic
luminosity ($L^{\star}_{{\rm X},0}$) are taken from the local XLF
determination which samples a median redshift of $z_0$.  This
redshift baseline is luminosity dependent as a result of the
flux-limited nature of the survey technique.  For instance in the BCS
sample, at \mbox{$L_{\rm X}$ = 0.1 (1, 10) $\times$ 10$^{44}$}
\lxunits, the median sample redshift is $z_0=0.02$ (0.06, 0.21).
Note that in the present model the faint-end slope of the luminosity
function ($\alpha$) is fixed at the local value since there is little
evidence for any change in this parameter as a function of redshift.

The sampling of the luminosity-redshift plane is sufficiently fine
that in each element, $dL_{\rm X}dz$, the number of observed clusters
is either zero or one and the expected number of clusters is
much smaller than unity.  In this sparse sampling limit we can define
a likelihood function $\cal{L}$ based on joint Poisson probabilities,

\begin{eqnarray}
{\cal{L}}~=~\prod_{i}\lambda(L_{{\rm X},i},z_{i})dL_{{\rm X}}dz~e^{-\lambda(L_{{\rm X},i},z_{i})dL_{{\rm X}}dz} \nonumber\\ \times~\prod_{j}e^{-\lambda(L_{{\rm X},j},z_{j})dL_{{\rm X}}dz .}
\label{eq:likelihood}
\end{eqnarray}

\noindent This is the combined probability of observing exactly one
cluster at each point ($L_{{\rm X},i},z_{i}$) populated by a 160SD
cluster and observing exactly zero clusters everywhere else ($L_{{\rm
X},j},z_{j}$).  Again, occupied elements in the luminosity-redshift
plane are indexed by $i$, whereas empty elements are indexed by $j$.
Transforming to the standard expression, $S=-2\ln{\cal{L}}$, and
dropping terms that are not model dependent, we have

\begin{eqnarray}
S = -2 \sum_{i} w_{i}~\ln[\lambda(L_{{\rm X},i},z_{i})]~\nonumber \\ +~2 \int_{L_{\rm X, min}}^{L_{\rm X, max}} \int_{z_{\rm min}}^{z_{\rm max}} \lambda(L_{{\rm X}},z) dL_{{\rm X}}dz .
\label{eq:S}
\end{eqnarray} 

\noindent A weighting term, $w_{i}$, is introduced to incorporate the
uncertainties in observed cluster luminosity and the summation is evaluated
over a total number of $i$ indices much greater than the number of
observed clusters \citep[e.g.,][]{Borgani2001b}.  Instead of being a
point in the observed ($L_{\rm X},z$)-plane, each cluster is smoothed
in the $L_{\rm X}$-direction according to a Gaussian distribution with
a width set by the 1$\sigma$ flux error, $\epsilon_{L_{\rm X}}$, for
each cluster (median value is 20\% for the 160SD sample).  No similar
treatment is required for the redshift measures since the typical
error is a few tenths of a percent.  Thus a weight is assigned to each
element in the luminosity-redshift plane based on the fractional
contributions of all clusters in the same redshift interval,
\begin{equation}
w_{i}~=~\sum_{k} \frac{1}{\sqrt{2\pi\epsilon_{L_{{\rm X},k}}^2}}~
                  \exp{\left[- 
                     \frac{(L_{{\rm X},k}-L_{{\rm X},i})^2}
                          {2\epsilon_{L_{{\rm X},k}}^2}
                   \right]} 
                   dL_{\rm X}.
\end{equation}

\noindent Here the summation $k$ is over the clusters with a redshift
between $z_{i}-dz/2$ and $z_{i}+dz/2$.

Recall the overall goal is to find the values of $A$ and $B$ that
predict a luminosity-redshift distribution that best matches the data.
These best-fit parameters are determined by minimizing $S$ with
confidence levels defined to be

\begin{equation}
\Delta S ~=~  S(A, B) - S(A_{\rm best}, B_{\rm best})
\end{equation}

\noindent Since $S$ is distributed like $\chi^2$, the $1\sigma$,
$2\sigma$, and $3\sigma$ (68.3\%, 95.4\%, and 99.7\%) confidence
intervals for a two parameter fit are $\Delta S = 2.30$, 6.17, and
11.8, respectively \citep{Avni1976,Cash1976,Cash1979}.

%................................................................
% FIGURE 8 --- IDL: /figs/misc.pro ---
\begin{figure*}
\epsscale{.4}
\plotone{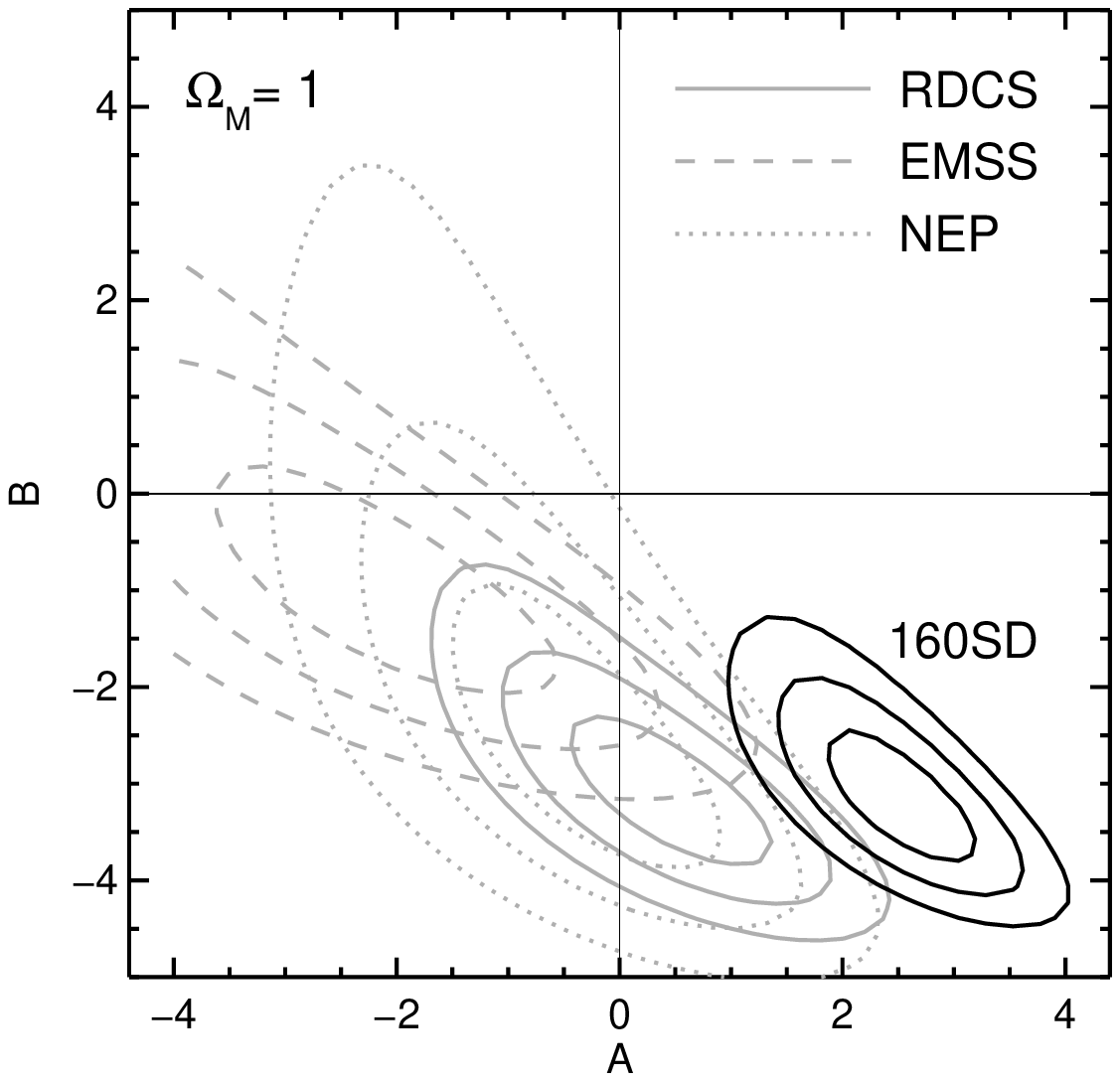}
\plotone{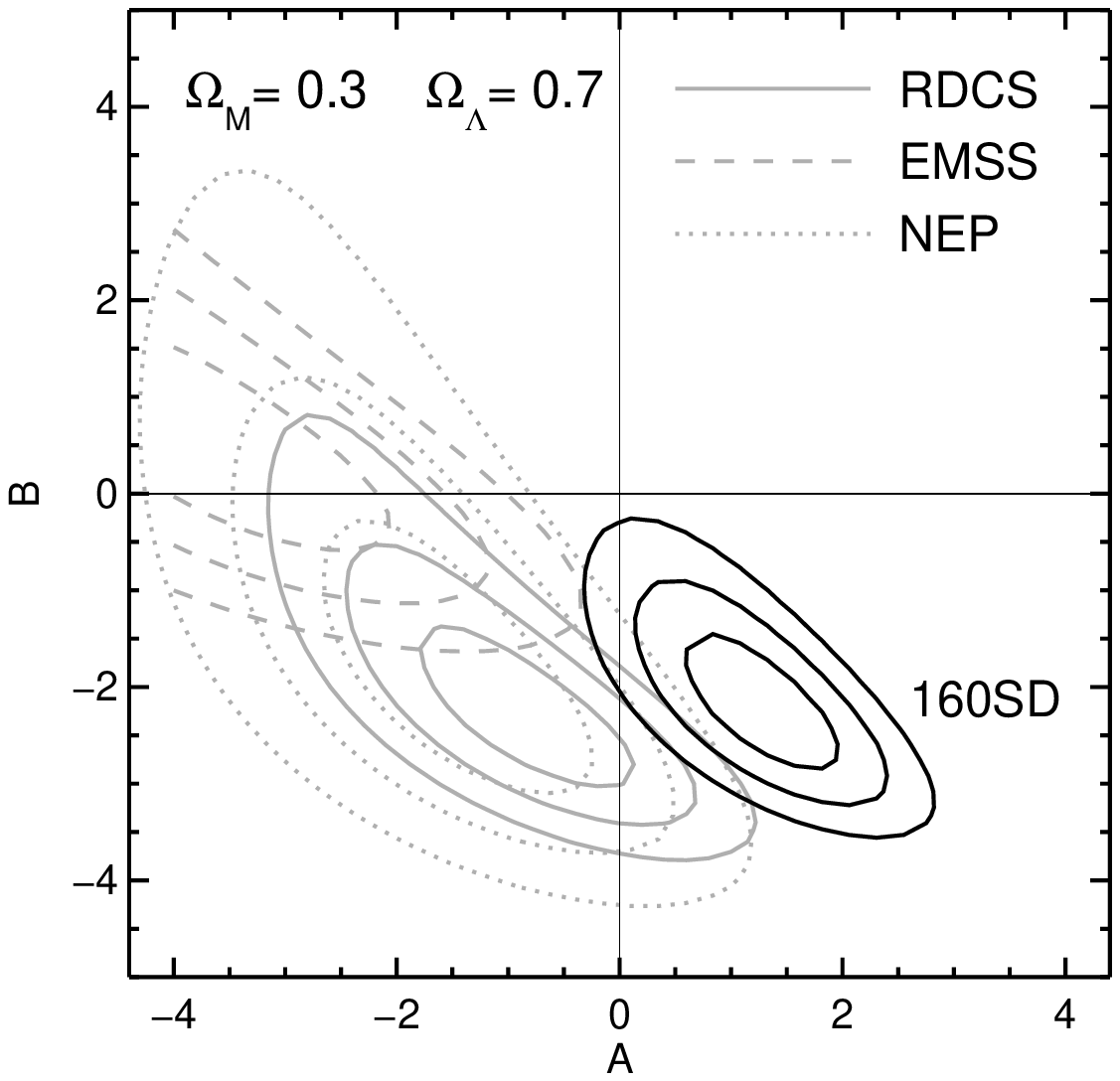}
\plotone{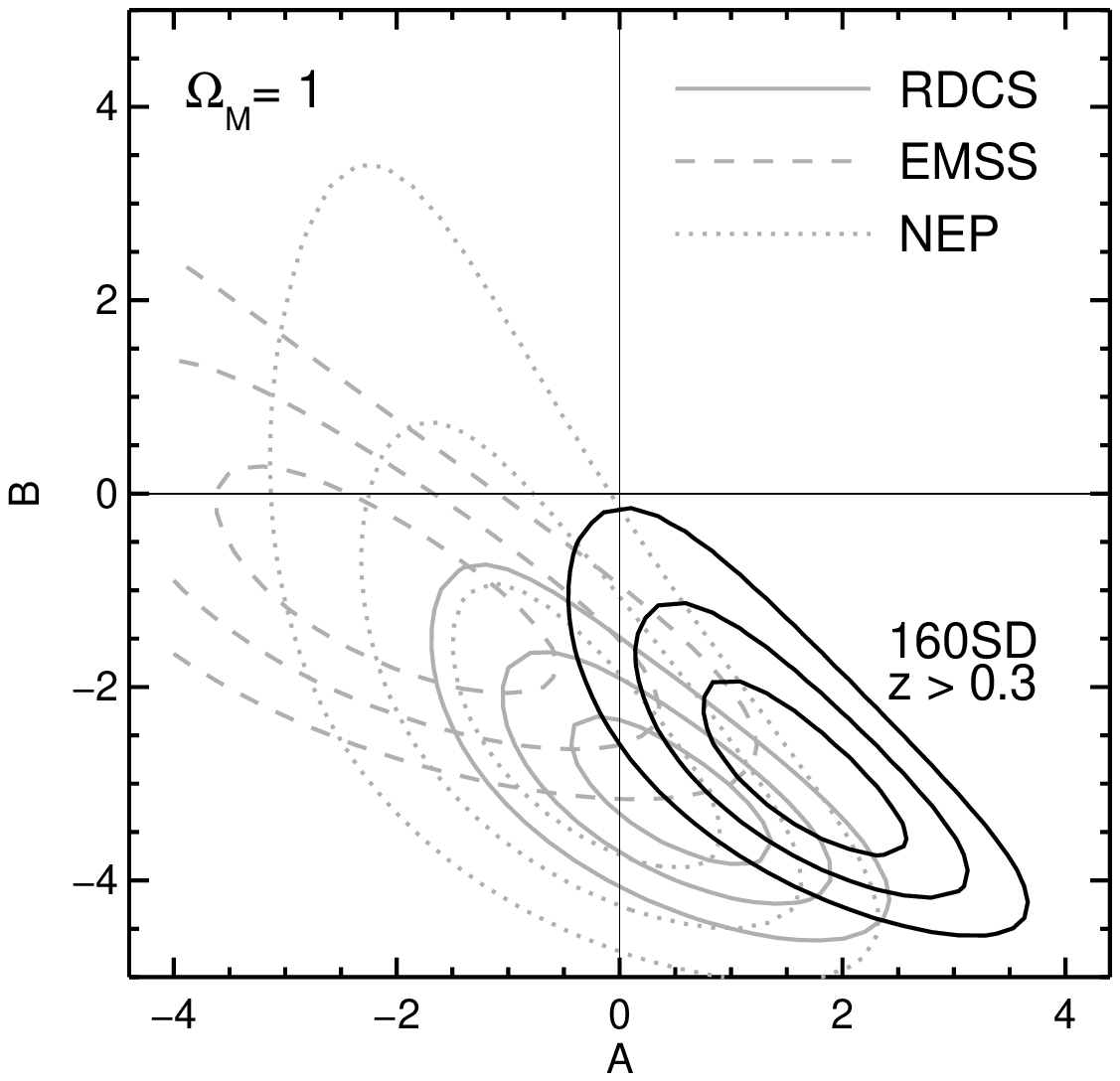}
\plotone{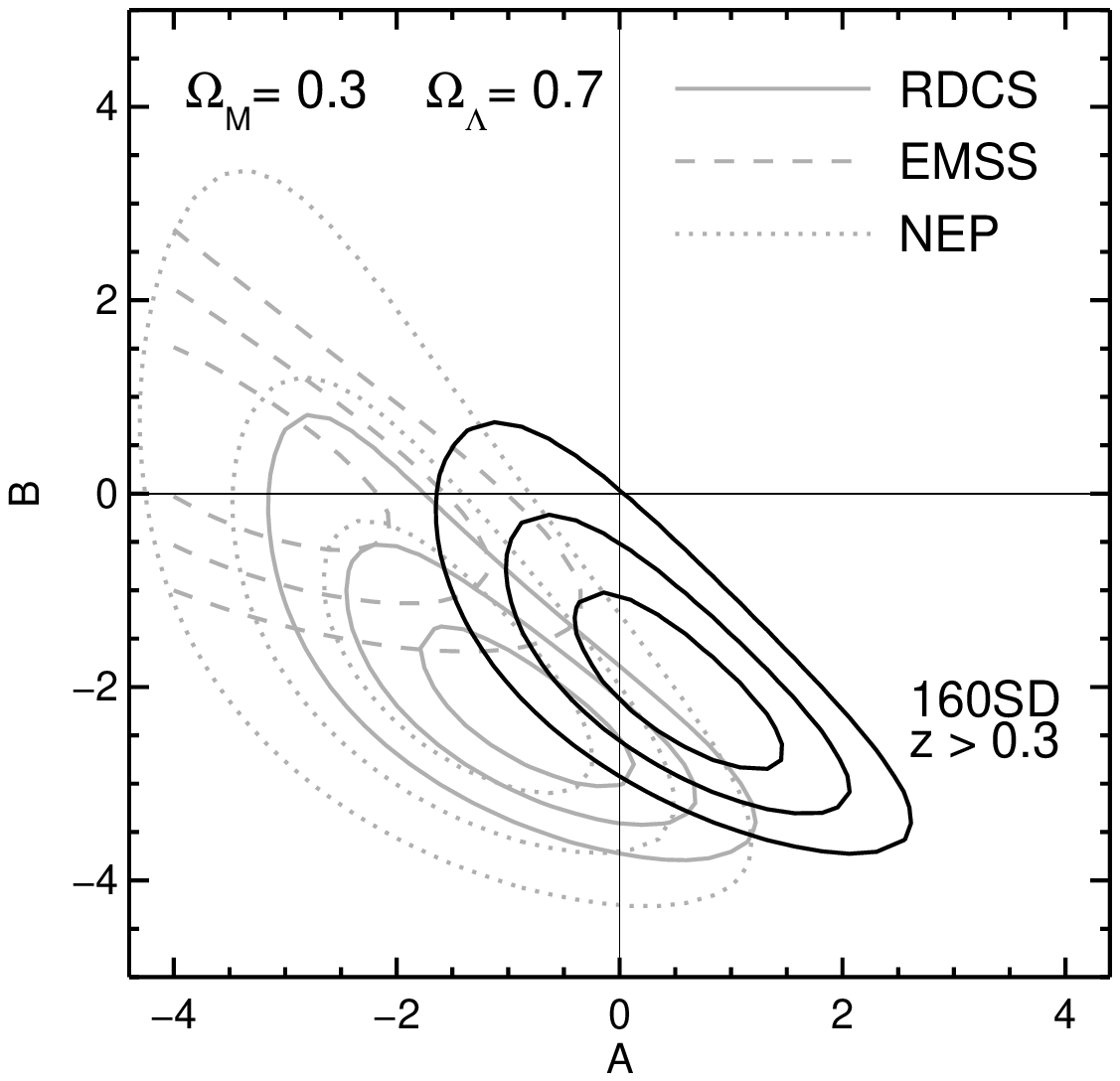}
\plotone{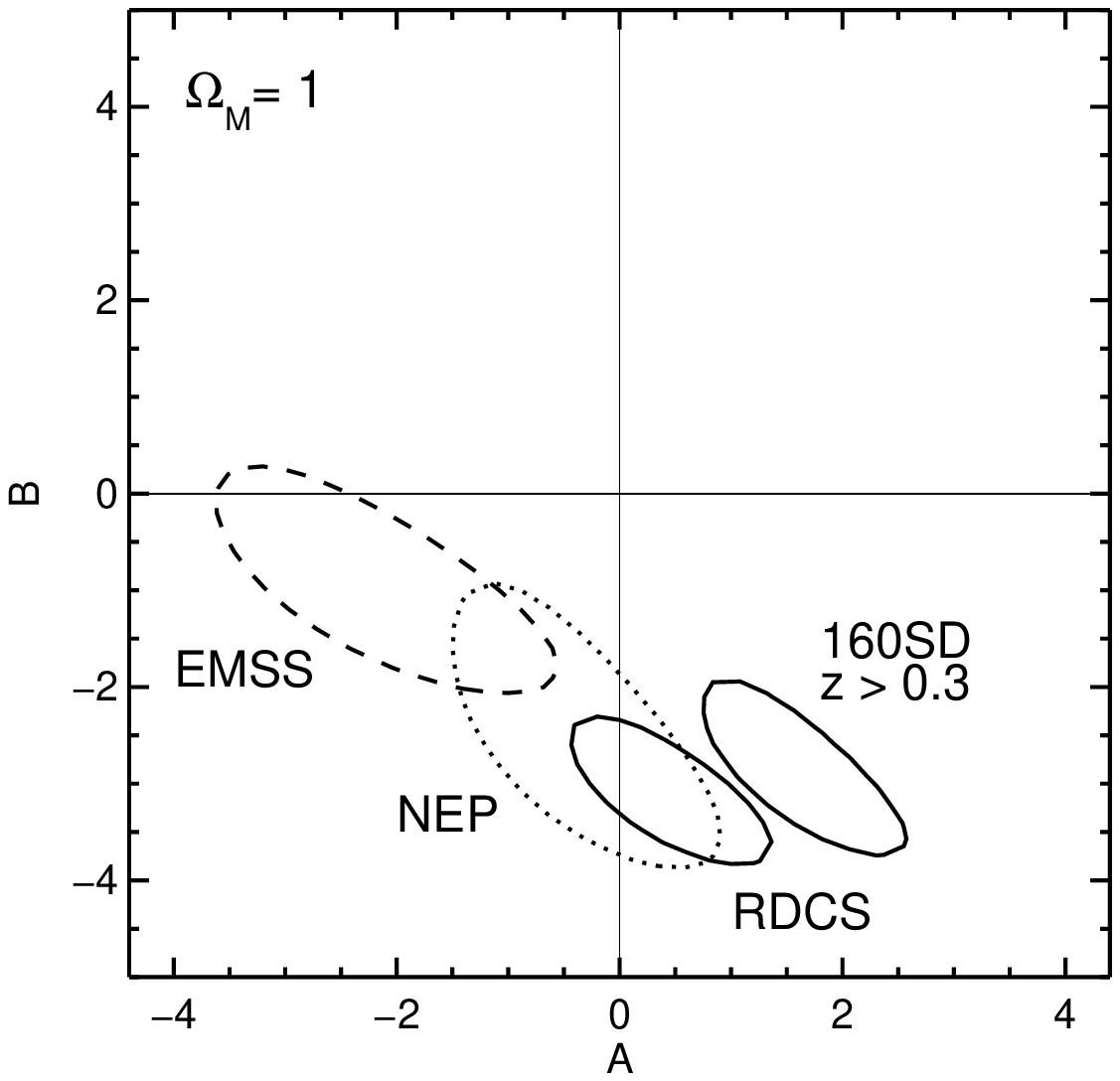}
\plotone{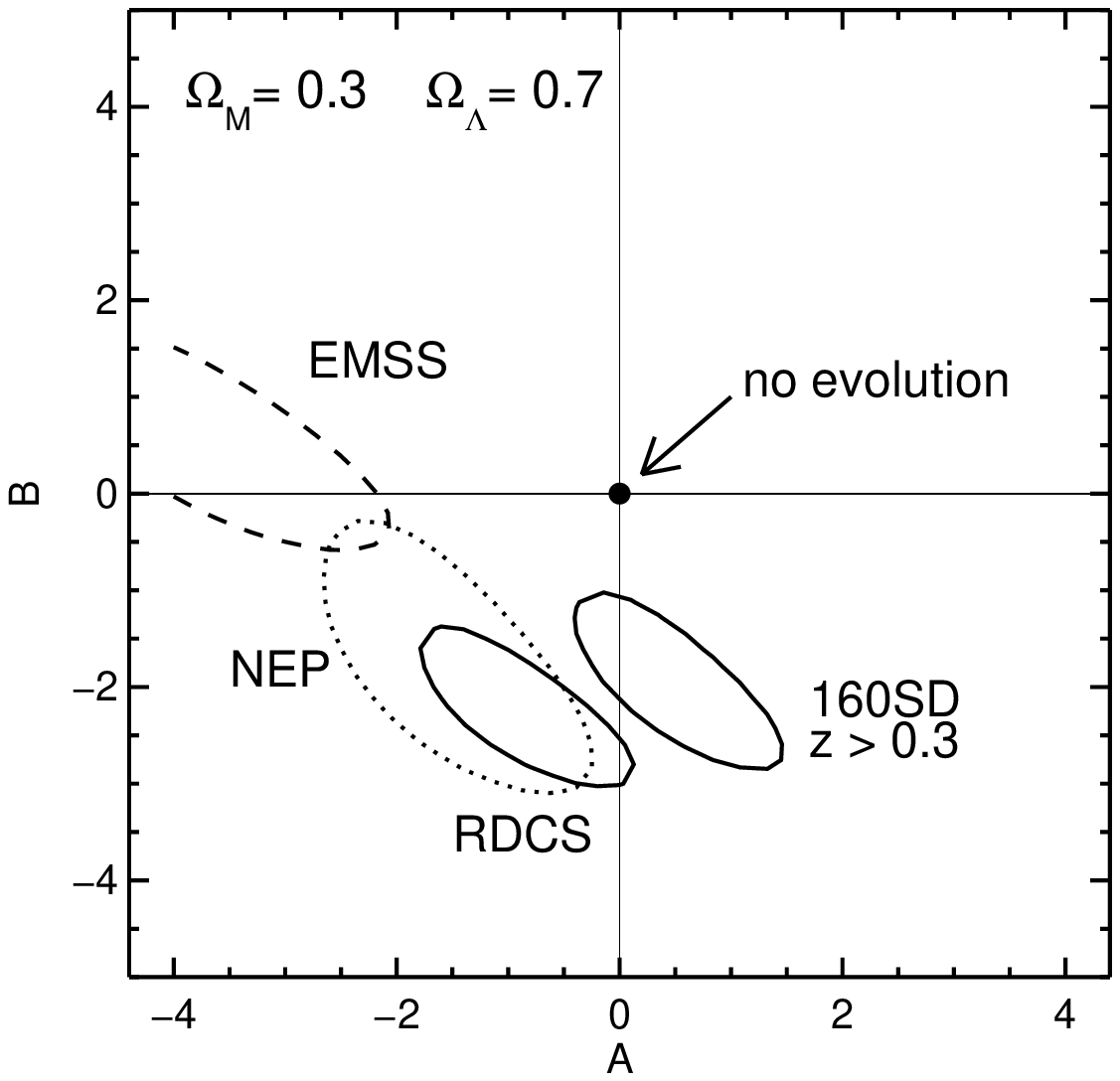}
\caption{Maximum-likelihood contours for the parameters A and B
characterizing the evolution of the cluster XLF where
\mbox{$\phi^{\star}\propto\phi^{\star}_{0}(1+z)^A$} and \mbox{$L_{\rm
X}^{\star} \propto L_{{\rm X},0}^{\star}(1+z)^B$} based on four
independent surveys.  Plotted are the results from our analysis of the
160SD and NEP samples, plus similar data for the EMSS and RDCS from
\citet{Rosati2002b}.  The no-evolution scenario corresponds to
$A=B=0$.  Confidence regions in the left column are computed in an
Einstein--de-Sitter universe, whereas those in the right column are
for a $\Lambda$-dominated universe.  TOP PANELS: One, two, and three
$\sigma$ contours with the 160SD results based on the entire sample.
MIDDLE PANELS: One, two, and three $\sigma$ contours with the 160SD
results based on the $z>0.3$ sample.  BOTTOM PANELS: Same as the
middle panels except only $1\sigma$ contours are shown for clarity.
\label{fig:ab}}
\end{figure*}
%................................................................

We use the maximum-likelihood procedure to estimate the evolutionary
parameters for a model XLF that best reproduces the observed cluster
luminosity-redshift distribution.  For the local parameters of the XLF
($\phi^{\star}_{0}$, $L_{{\rm X},0}$, and $\alpha$) we adopt the
values from the BCS.  This permits comparisons to previous work and
assumes a conservative position since the BCS has the lowest
normalization of the three local references.  We estimated the
baseline redshift, $z_0(L_{\rm X}$), by calculating the median
redshifts in seven luminosity intervals of the BCS sample and
interpolating across these points ($z_0$ = 0.018, 0.028, 0.040, 0.061,
0.103, 0.171, 0.236 at $L_{\rm X,EdS}$/10$^{44}$ = 0.054, 0.139, 0.303,
0.938, 2.283, 5.771, 12.929 \lxunits).  The top panels of \mbox{Figure
\ref{fig:ab}} indicate the constraints using the entire 160SD sample
of 177 clusters ($0.02 < z < 0.80$, 10$^{42} < L_{\rm X} < 10^{46}$,
$f_{\rm X} > 4 \times 10^{-14}$) in two different cosmologies:
\mbox{$A=2.6^{+0.6}_{-0.7}$}, \mbox{$B=-3.2^{+0.8}_{-0.6}$} (EdS),
\mbox{$A=1.3^{+0.6}_{-0.7}$}, \mbox{$B=-2.3^{+0.8}_{-0.6}$}
($\Lambda$-cosmo).  In either case, no evolution \mbox{($A=B=0$)} is
strongly excluded at $>3\sigma$.

To examine the evolution ``signal'' in the 160SD sample at higher
redshifts corresponding to the XLFs in \mbox{Figures \ref{fig:hizxlf}
\& \ref{fig:hizxlf_lcdm}}, we repeat the maximum-likelihood analysis
using the 66 clusters at $z>0.3$.  The resulting contours are shown in
the middle panels of \mbox{Figure \ref{fig:ab}} and the best-fit
values are: \mbox{$A=1.6^{+1.0}_{-0.8}$},
\mbox{$B=-2.9^{+1.0}_{-0.8}$} (EdS), \mbox{$A=0.6^{+0.9}_{-1.0}$},
\mbox{$B=-2.1^{+1.0}_{-0.8}$} ($\Lambda$-cosmo).  Relative to the full
sample analysis, the best-fit $A$ is about $1\sigma$ lower while the
best-fit $B$ is effectively unchanged, being a few tenths of
$\sigma$ lower.  Apparently, the $A\/$ parameter in the full analysis
is elevated to fit an enhancement of clusters at $z<0.3$ which can be
seen at \mbox{$\sim3\times$ 10$^{43}$} \lxunits~in the low-redshift
XLF.  We adopt the less-biased $z>0.3$ results as representative of
the 160SD in subsequent discussions.

For comparison purposes, we also show in \mbox{Figure \ref{fig:ab}}
the constraints from three additional datasets --- the NEP, EMSS, and
RDCS.  One, two and three $\sigma$ contours are shown in the top and
middle panels, whereas only $1\sigma$ contours are plotted in
bottom panels.  The degree of agreement amongst the four surveys is
reasonably good but certainly not perfect.  It is not immediately
clear how much of this is due to potential systematics in the
individual datasets or the appropriateness of the model XLF used in
the maximum-likelihood fit.  These issues will be addressed in a
forthcoming paper focused on the joint analysis of the available
samples \citep[J.\ P.\@ Henry et al. 2004, in preparation; preliminary
results discussed in][]{Henry2003}.

\begin{figure}
\epsscale{1.15}
\plotone{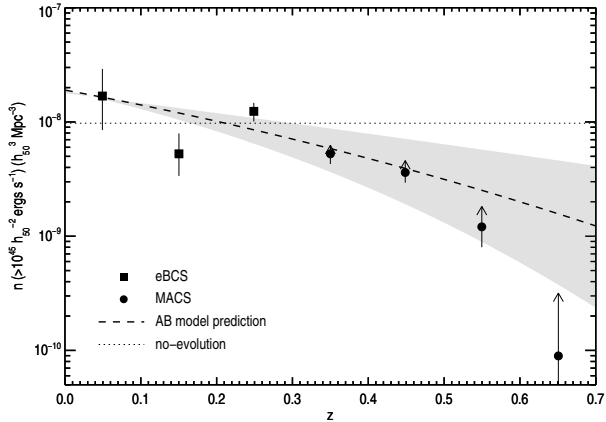}
\caption{Comoving volume density of very X-ray luminous clusters with
$L_{\rm X} > 10^{45}$ \lxunits~ (0.1--2.4\,keV; equivalently
\mbox{$>$6.2 $\times$ 10$^{44}$}, 0.5--2.0\,keV).  The data points are
derived from the eBCS and the MACS samples as reported by H. Ebeling et al. \citep[see Figure 5 in][and H. Ebeling et al. 2004, in preparation]{Ellis2002b} in an Einstein--de-Sitter
universe. The dashed line and shaded region indicate the prediction
based on our ensemble best-fit model XLF ($A=0, B=-2.5\pm1$); {\em
not} a fit to the eBCS+MACS data. The dotted line marks the
no-evolution expectation derived from the BCS Schechter function fit to the local XLF which is dominated by $z\sim0.2$ clusters at these luminosities. \label{fig:1045}}
\end{figure}

Considering the ensemble of data from four independent cluster
samples, the no-evolution scenario \mbox{($A=B=0$)} is strongly ruled
out at $\gg3\sigma$ at $z\la 0.8$ regardless of cosmology.  In an
Einstein--de-Sitter universe, the change in the cluster population is
consistent with pure luminosity evolution ($A\approx 0, B\approx
-2.5$), whereas it may be a combination of luminosity and density
evolution ($A\approx -1, B\approx -2$) in a $\Lambda$-dominated
universe.  Given the lack of clusters with $L_{\rm X} > 10^{45}$
\lxunits~in the analyzed samples, caution is warranted if these
results are extrapolated to predict the cluster abundance at very high
luminosities.  Nonetheless, we show in \mbox{Figure \ref{fig:1045}}
that our characterization of cluster evolution is consistent with a
direct measurement of the comoving volume density of clusters at
$L_{\rm X} > 10^{45}$ from the eBCS+MACS surveys (H. Ebeling et
al. 2004, in preparation).

%................................................................
% FIGURE 9 --- IDL: /figs/fig9.pro ---
%................................................................

%................................................................
% FIGURE 10 --- IDL: /figs/fig10.pro ---
\begin{figure*}
\epsscale{.95}
\plotone{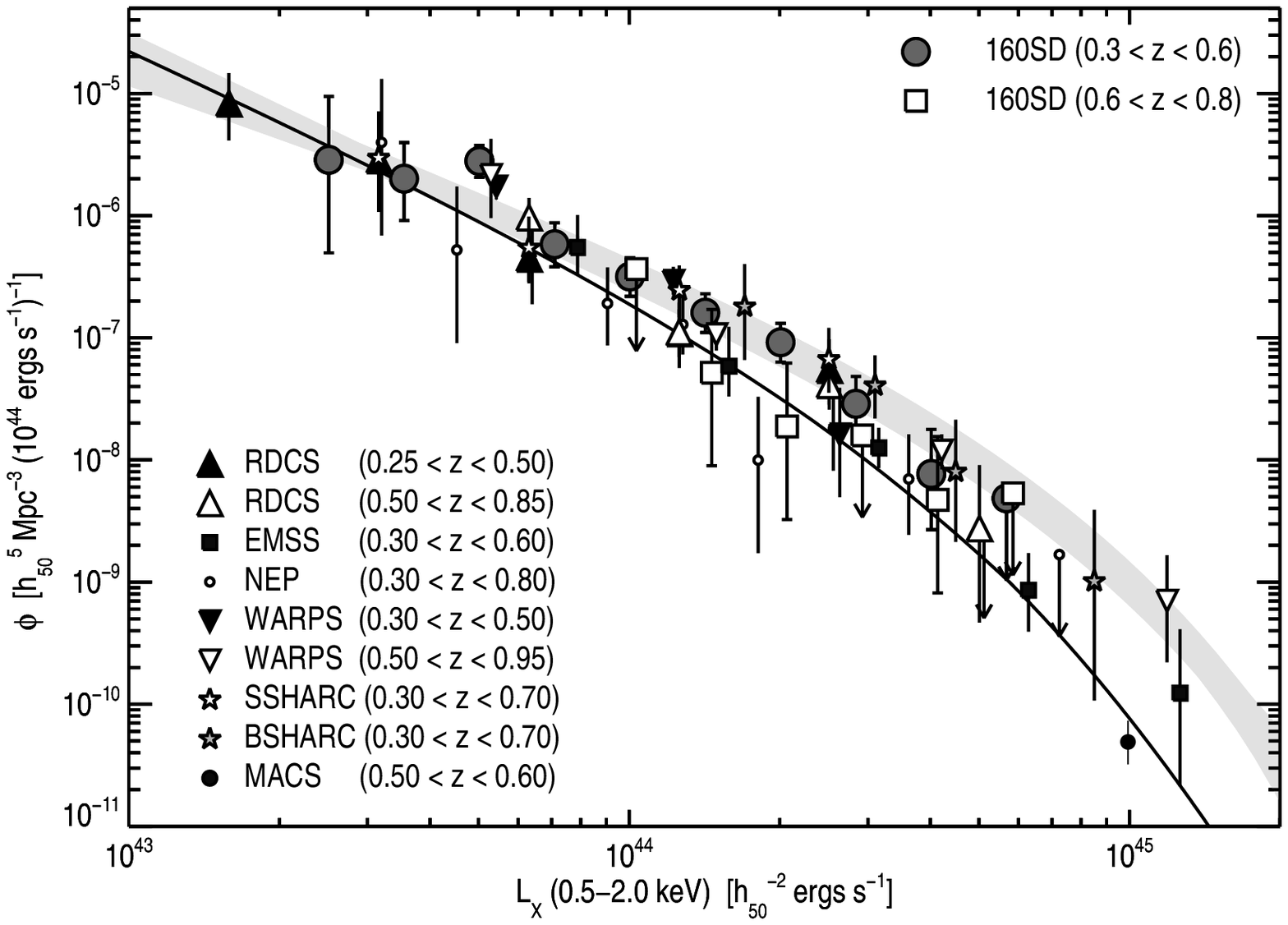}
\caption{Compilation of high-redshift XLFs as measured by eight
independent X-ray flux-limited surveys (Einstein--de-Sitter universe).
References are 160SD (this paper), RDCS \citep{Rosati2000}, EMSS
\citep{Henry1992}, NEP \citep{Gioia2001}, WARPS \citep{Jones2000c},
SSHARC \citep{Burke1997}, BSHARC \citep{Nichol1999}, and MACS
(H. Ebeling 2003, private communication). The shaded region delineates
the regime of the local XLF whereas the heavy solid line is an
evolving model XLF ($A=0$, $B=-2.5$) at $z=0.7$.\label{fig:hizxlfb}}
\end{figure*}
%................................................................

%--------------- SECTION 6: CONCLUSIONS ------------------
\section{Conclusions}
\label{Conclusions}

We have used the 160SD sample of 201 X-ray selected galaxy clusters to
track the volume density of these systems at local, intermediate and
high redshifts.  Nonparametric measurements of the cluster XLF
suggest there is effectively no detectable evolution in the population
out to $z\sim 0.6$ at luminosities less than a few times
\mbox{10$^{44}$ $h_{50}^{-2}$} \lxunits.  However, data in the
redshift interval $0.6 < z < 0.8$ indicate a mild but significant
($>3\sigma$) change in volume densities above luminosities of
approximately \mbox{10$^{44}$ $h_{50}^{-2}$} \lxunits.  Our findings
demonstrate a deficit of high-luminosity clusters at high redshift
relative to the present-day levels.  For example, we observe only 3
clusters at $0.6 < z < 0.8$ where an integral over the local XLF
predicts we should find at the very least 16 clusters.  A
maximum-likelihood analysis of the observed luminosity-redshift
distribution further underscores that the no-evolution scenario is
entirely inconsistent with 160SD data.  Modelling the XLF with an
evolving Schechter function, we have demonstrated that the 160SD, NEP,
RDCS and EMSS clusters samples all reject a model lacking evolution.
Further our evolving XLF model derived from datasets at $L_{\rm X} <
10^{45}$ successfully predicts the observed volume densities at $L_{\rm
X} > 10^{45}$.

A composite view of the high-redshift cluster population is shown in
\mbox{Figure \ref{fig:hizxlfb}}.  Here we plot the latest compilation
of distant XLFs as measured by eight X-ray flux-limited surveys.  From
these data six independent teams have concluded there is measurable
cluster evolution --- EMSS: \citet{Gioia1990a,Henry1992}, BSHARC:
\citet{Nichol1999}, RDCS: \citet{Rosati2000}, NEP: \citet{Gioia2001},
160SD: this work, and MACS (H. Ebeling 2004, in preparation).
Conversely, SSHARC \citep{Burke1997} and WARPS \citep{Jones2000} draw
the opposite conclusion.  Note however that the sky coverage of the
SSHARC survey (17.7 deg$^{2}$) is inadequate to measure evolutionary
changes at the very bright end of the XLF, as studied in the large
solid angle EMSS (735 deg$^{2}$) and the 160SD surveys.

Measurements of the cluster X-ray luminosity function are consistent
with a population whose comoving volume density evolves as a function
of both redshift and luminosity (mass).  For a sample of distant
clusters in a fixed luminosity interval, the difference in volume
density for this sample relative to the present-day population
increases with increasing redshift (e.g., \mbox{Figure
\ref{fig:1045}}).  Considering a sample of clusters in a distant
redshift interval, the difference in volume density for this sample
and the present-day population increases with increasing luminosity
(e.g., \mbox{Figure \ref{fig:hizxlfb}}).  Visualizing this on the
luminosity-redshift plane, the cluster deficit is maximal at high
redshift and high luminosity (i.e., the upper-right corner of Figure
\ref{fig:lxzall}).  The maximum-likelihood results in \mbox{Figure
\ref{fig:ab}} demonstrate that the EMSS, 160SD, RDCS, and NEP surveys
all reject the no-evolution scenario and generally agree on the
character of the evolution (e.g., AB contours overlap at 1--2
$\sigma$).  The EMSS was able to detect a statistically significant
cluster deficit in the \mbox{$0.3 < z < 0.6$} redshift shell because
of their sensitivity to very luminous clusters.  Surveys such as the
160SD probe lower luminosities where the evolution ``signal'' is
intrinsically weaker and thus have to push to higher redshifts (e.g.,
\mbox{$0.6 < z < 0.8$}) before the deficit grows large enough for a
significant detection.

The preponderance of observational data requires mild evolution in the
bright end of the cluster XLF as measured out to $z\sim 0.8$, whereas
the volume densities of the bulk of the population (low and
intermediately massive systems) do not change appreciably.  In the context
of hierarchical structure formation, we are probing sufficiently large
mass aggregations at sufficiently early times where the Universe has
yet to assemble these clusters to present-day volume densities.
Cosmological constraints derived from the RDCS and 160SD samples
\citep{Borgani2001b,Vikhlinin2003} demonstrate the observed evolution
is consistent with structure growth in a flat universe with
$\Omega_{\rm M} \sim 0.3$.

A decade after the first evidence for negative cluster evolution
reported by the {\em Einstein} EMSS survey
\citep{Gioia1990a,Henry1992}, we are now witnessing the fruition of
the {\em ROSAT\/} era cluster surveys: 1) the robust determination of
the local cluster X-ray luminosity function, and 2) multiple,
independent confirmations of the EMSS results on cluster evolution.
Further work remains to assimilate the available cluster samples in a
detailed, joint analysis to produce maximal constraints on the
evolution phenomenon.  Furthermore, additional inputs are expected in
the near term from two on-going surveys --- BMW-HRI
\citep{Panzera2003} and MACS \citep{Ebeling2001c}.  The BMW-HRI has a
sky coverage rather similar to the 160SD, and hence will likely
provide similar XLF measures.  The final results from MACS should
prove particularly interesting because this program probes extremely
high luminosities \mbox{($L_{\rm X} \ga 10^{45}$} \lxunits) at high
redshift exactly where the evolution signature should be strongest.
Finally, on longer timescales, searches with {\em XMM-Newton\/} and
possibly a new dedicated X-ray survey satellite should reveal
substantial numbers of $z > 1$ clusters which will provide powerful
leverage in the study of cluster evolution.

\acknowledgements 

It is a pleasure to thank Piero Rosati, Colin Norman, Stefano Borgani
and Emanuele Daddi for helpful discussions during the course of this
work.  Hans B\"{o}hringer, Harald Ebeling, Sabrina DeGrandi, and Piero
Rosati kindly provided supporting information concerning their
respective XLF measurements.  We thank the referee for comments that
improved the presentation of this work.

C.R.M. acknowledges financial support from the ESO Office for Science.
The contributions of B.R.M., A.V., W.R.F., and C.J.F. were possible
thanks in part to NASA grant NAG5-9217 and contract NAS8-39073.  HQ
was partially supported by the Guggenheim Foundation and FONDAP Centro
de Astrofisica.
%\newpage

\bibliographystyle{apj}
\bibliography{myrefs}

\end{document}

%% file: tab1.tex
\begin{deluxetable*}{lclll}
\tablewidth{0pt} 

\tablecaption{Best-Fit Schechter Parameters for the Local Cluster XLF ($z < 0.3$) \label{tab:schechter}}
\tablehead{\colhead{Sample} & \colhead{$L_{\rm X}^{\star}$\tablenotemark{\dag}} & \colhead{$\alpha$} & \colhead{$\phi^{\star}$} & \colhead{Reference} \\
           \colhead{} & \colhead{($10^{44}$ $h_{50}^{-2}$ ergs s$^{-1}$)} & \colhead{} & \colhead{$h_{50}^{3}$ Mpc$^{-3}$}} 

\startdata
\\
BCS     & $5.70^{+1.29}_{-0.93}$ & $1.85^{+0.09}_{-0.09}$ & ($7.56^{+0.82}_{-0.75}$) $\times$ 10$^{-8}$ & \citet{Ebeling1997} \\
\\
RASS1BS & $3.80^{+0.70}_{-0.55}$ & $1.52^{+0.11}_{-0.11}$ & ($2.53^{+0.22}_{-0.22}$) $\times$ 10$^{-7}$ & \citet{DeGrandi1999a} \\
\\
REFLEX\tablenotemark{\ddag}  & $4.21^{+0.37}_{-0.34}$ & $1.63^{+0.06}_{-0.06}$ & ($1.80^{+0.5\phantom{0}}_{-0.4\phantom{0}}$) $\times$ 10$^{-7}$  & \citet{Boehringer2002} \\
  & $5.18^{+0.56}_{-0.50}$ & $1.69^{+0.045}_{-0.045}$ & \phantom{(}$1.07^{\phantom{+0.50}}_{\phantom{-0.40}}$\phantom{)} $\times$ 10$^{-7}$  & $\Omega_{M}=0.3$, $\Omega_{\Lambda} = 0.7$ \\
\\
\enddata
\tablecomments{Parameters for an Einstein--de-Sitter universe unless otherwise indicated}
\tablenotetext{\dag}{$L_{\rm X}^{\star}$ is quoted in the 0.5--2.0 keV band}
\tablenotetext{\ddag}{fit includes a correction for missing flux for consistency with the other surveys \citep[see details in][]{Boehringer2002}}

\end{deluxetable*}

%% file: tab2.tex
\begin{deluxetable*}{cccccccc}
\tablewidth{0pt} 
\tablecaption{Local 160SD Cluster X-ray Luminosity Function ($0.02 < z < 0.3$, EdS)  \label{tab:lozxlf}}

\tablehead{\colhead{$L_{\rm X}$(center)\tablenotemark{\dag}} &
           \colhead{$L_{\rm X}$(min)    \tablenotemark{\dag}} &
           \colhead{$L_{\rm X}$(max)    \tablenotemark{\dag}} &
           \colhead{$\phi(L_{\rm X})$} &
           \colhead{$\phi(L_{\rm X})_{-1\sigma}$} &
           \colhead{$\phi(L_{\rm X})_{+1\sigma}$} &
           \colhead{$<z>$} &
           \colhead{$N_{\rm cl}$ } \\
           \colhead{($10^{44}$ $h_{50}^{-2}$ ergs s$^{-1}$)} &
           \colhead{$\cdots$} &
           \colhead{$\cdots$} &
           \colhead{($h_{50}^{5}$ Mpc$^{-3}$ (10$^{44}$ ergs s$^{-1}$)$^{-1}$)} &
           \colhead{$\cdots$} &
           \colhead{$\cdots$} }

\startdata
\\
% Produced by xlf2latex.pro from call in fig45-localxlf.pro
\input{tab2-data.tex}
\enddata
\tablenotetext{\dag}{$L_{\rm X}$ is quoted in the 0.5--2.0 keV band}

\end{deluxetable*}

%% file: tab2-data.tex
 0.016 &  0.013 &  0.019 & 5.09 $\times$ 10$^{-4}$ & 1.78 $\times$ 10$^{-4}$ & 1.18 $\times$ 10$^{-3}$ & 0.045 &  2 \\
 0.022 &  0.019 &  0.026 & 2.19 $\times$ 10$^{-4}$ & 7.65 $\times$ 10$^{-5}$ & 5.07 $\times$ 10$^{-4}$ & 0.035 &  2 \\
 0.031 &  0.026 &  0.037 & 1.42 $\times$ 10$^{-4}$ & 6.45 $\times$ 10$^{-5}$ & 2.81 $\times$ 10$^{-4}$ & 0.062 &  3 \\
 0.044 &  0.037 &  0.053 & 2.08 $\times$ 10$^{-5}$ & 3.59 $\times$ 10$^{-6}$ & 6.85 $\times$ 10$^{-5}$ & 0.135 &  1 \\
 0.063 &  0.053 &  0.075 & 4.59 $\times$ 10$^{-5}$ & 2.61 $\times$ 10$^{-5}$ & 7.69 $\times$ 10$^{-5}$ & 0.142 &  5 \\
 0.089 &  0.075 &  0.106 & 2.87 $\times$ 10$^{-5}$ & 1.80 $\times$ 10$^{-5}$ & 4.42 $\times$ 10$^{-5}$ & 0.136 &  7 \\
 0.125 &  0.106 &  0.149 & 1.86 $\times$ 10$^{-5}$ & 1.28 $\times$ 10$^{-5}$ & 2.65 $\times$ 10$^{-5}$ & 0.156 & 10 \\
 0.177 &  0.149 &  0.211 & 6.98 $\times$ 10$^{-6}$ & 4.55 $\times$ 10$^{-6}$ & 1.04 $\times$ 10$^{-5}$ & 0.154 &  8 \\
 0.251 &  0.211 &  0.298 & 7.44 $\times$ 10$^{-6}$ & 5.65 $\times$ 10$^{-6}$ & 9.72 $\times$ 10$^{-6}$ & 0.200 & 17 \\
 0.355 &  0.298 &  0.422 & 5.49 $\times$ 10$^{-6}$ & 4.35 $\times$ 10$^{-6}$ & 6.89 $\times$ 10$^{-6}$ & 0.197 & 23 \\
 0.502 &  0.422 &  0.597 & 1.27 $\times$ 10$^{-6}$ & 8.55 $\times$ 10$^{-7}$ & 1.86 $\times$ 10$^{-6}$ & 0.195 &  9 \\
 0.710 &  0.597 &  0.844 & 9.89 $\times$ 10$^{-7}$ & 6.94 $\times$ 10$^{-7}$ & 1.39 $\times$ 10$^{-6}$ & 0.239 & 11 \\
 1.004 &  0.844 &  1.194 & 3.60 $\times$ 10$^{-7}$ & 2.16 $\times$ 10$^{-7}$ & 5.76 $\times$ 10$^{-7}$ & 0.226 &  6 \\
 1.420 &  1.194 &  1.688 & 8.26 $\times$ 10$^{-8}$ & 2.89 $\times$ 10$^{-8}$ & 1.92 $\times$ 10$^{-7}$ & 0.225 &  2 \\
 2.008 &  1.688 &  2.388 & 5.76 $\times$ 10$^{-8}$ & 2.01 $\times$ 10$^{-8}$ & 1.34 $\times$ 10$^{-7}$ & 0.180 &  2 \\
 2.840 &  2.388 &  3.377 & 2.01 $\times$ 10$^{-8}$ & 3.47 $\times$ 10$^{-9}$ & 6.63 $\times$ 10$^{-8}$ & 0.296 &  1 \\

%% file: tab3.tex
\begin{deluxetable*}{cccccccc}
\tablewidth{0pt} 

\tablecaption{Distant 160SD Cluster X-ray Luminosity Function (EdS)  \label{tab:xlf}}

\tablehead{\colhead{$L_{\rm X}$(center)\tablenotemark{\dag}} &
           \colhead{$L_{\rm X}$(min)    \tablenotemark{\dag}} &
           \colhead{$L_{\rm X}$(max)    \tablenotemark{\dag}} &
           \colhead{$\phi(L_{\rm X})$} &
           \colhead{$\phi(L_{\rm X})_{-1\sigma}$} &
           \colhead{$\phi(L_{\rm X})_{+1\sigma}$} &
           \colhead{$<z>$} &
           \colhead{$N_{\rm cl}$ } \\
           \colhead{($10^{44}$ $h_{50}^{-2}$ ergs s$^{-1}$)} &
           \colhead{$\cdots$} &
           \colhead{$\cdots$} &
           \colhead{($h_{50}^{5}$ Mpc$^{-3}$ (10$^{44}$ ergs s$^{-1}$)$^{-1}$)} &
           \colhead{$\cdots$} &
           \colhead{$\cdots$} }
\startdata
\\
\multicolumn{1}{c}{$0.3 < z < 0.6$}\\
\input{tab3a-data.tex}
\\
\multicolumn{1}{c}{$0.6 < z < 0.8$}\\
\input{tab3b-data.tex}
\\
\enddata
\tablenotetext{\dag}{$L_{\rm X}$ is quoted in the 0.5--2.0 keV band}
\end{deluxetable*}

%% file: tab3a-data.tex
 0.251 &  0.211 &  0.298 & 2.86 $\times$ 10$^{-6}$ & 4.95 $\times$ 10$^{-7}$ & 9.45 $\times$ 10$^{-6}$ & 0.329 &  1 \\
 0.355 &  0.298 &  0.422 & 2.01 $\times$ 10$^{-6}$ & 9.10 $\times$ 10$^{-7}$ & 3.96 $\times$ 10$^{-6}$ & 0.334 &  3 \\
 0.502 &  0.422 &  0.597 & 2.80 $\times$ 10$^{-6}$ & 2.06 $\times$ 10$^{-6}$ & 3.77 $\times$ 10$^{-6}$ & 0.376 & 14 \\
 0.710 &  0.597 &  0.844 & 5.83 $\times$ 10$^{-7}$ & 3.80 $\times$ 10$^{-7}$ & 8.72 $\times$ 10$^{-7}$ & 0.420 &  8 \\
 1.004 &  0.844 &  1.194 & 3.17 $\times$ 10$^{-7}$ & 2.18 $\times$ 10$^{-7}$ & 4.53 $\times$ 10$^{-7}$ & 0.461 & 10 \\
 1.420 &  1.194 &  1.688 & 1.61 $\times$ 10$^{-7}$ & 1.10 $\times$ 10$^{-7}$ & 2.29 $\times$ 10$^{-7}$ & 0.418 & 10 \\
 2.008 &  1.688 &  2.388 & 9.18 $\times$ 10$^{-8}$ & 6.32 $\times$ 10$^{-8}$ & 1.31 $\times$ 10$^{-7}$ & 0.517 & 10 \\
 2.840 &  2.388 &  3.377 & 2.88 $\times$ 10$^{-8}$ & 1.64 $\times$ 10$^{-8}$ & 4.83 $\times$ 10$^{-8}$ & 0.501 &  5 \\
 4.016 &  3.377 &  4.776 & 7.67 $\times$ 10$^{-9}$ & 2.68 $\times$ 10$^{-9}$ & 1.78 $\times$ 10$^{-8}$ & 0.485 &  2 \\
 5.679 &  4.776 &  6.754 & 4.85 $\times$ 10$^{-9}$ & $\cdots$ & 4.85 $\times$ 10$^{-9}$ & $\cdots$ &  0 \\

%% file: tab3b-data.tex
 1.004 &  0.844 &  1.194 & 3.65 $\times$ 10$^{-7}$ & $\cdots$ & 3.65 $\times$ 10$^{-7}$ & $\cdots$ &  0 \\
 1.420 &  1.194 &  1.688 & 5.17 $\times$ 10$^{-8}$ & 8.95 $\times$ 10$^{-9}$ & 1.71 $\times$ 10$^{-7}$ & 0.699 &  1 \\
 2.008 &  1.688 &  2.388 & 1.88 $\times$ 10$^{-8}$ & 3.25 $\times$ 10$^{-9}$ & 6.20 $\times$ 10$^{-8}$ & 0.625 &  1 \\
 2.840 &  2.388 &  3.377 & 1.59 $\times$ 10$^{-8}$ & $\cdots$ & 1.59 $\times$ 10$^{-8}$ & $\cdots$ &  0 \\
 4.016 &  3.377 &  4.776 & 4.71 $\times$ 10$^{-9}$ & 8.15 $\times$ 10$^{-10}$ & 1.55 $\times$ 10$^{-8}$ & 0.700 &  1 \\
 5.679 &  4.776 &  6.754 & 5.33 $\times$ 10$^{-9}$ & $\cdots$ & 5.33 $\times$ 10$^{-9}$ & $\cdots$ &  0 \\

%% file: tab4.tex
\begin{deluxetable*}{cccccccc}
\tablewidth{0pt} 

\tablecaption{Distant 160SD Cluster X-ray Luminosity Function ($\Omega_{M}=0.3$ and $\Omega_{\Lambda} = 0.7$)  \label{tab:xlf_lcdm}}

\tablehead{\colhead{$L_{\rm X}$(center)\tablenotemark{\dag}} &
           \colhead{$L_{\rm X}$(min)    \tablenotemark{\dag}} &
           \colhead{$L_{\rm X}$(max)    \tablenotemark{\dag}} &
           \colhead{$\phi(L_{\rm X})$} &
           \colhead{$\phi(L_{\rm X})_{-1\sigma}$} &
           \colhead{$\phi(L_{\rm X})_{+1\sigma}$} &
           \colhead{$<z>$} &
           \colhead{$N_{\rm cl}$ } \\
           \colhead{($10^{44}$ $h_{50}^{-2}$ ergs s$^{-1}$)} &
           \colhead{$\cdots$} &
           \colhead{$\cdots$} &
           \colhead{($h_{50}^{5}$ Mpc$^{-3}$ (10$^{44}$ ergs s$^{-1}$)$^{-1}$)} &
           \colhead{$\cdots$} &
           \colhead{$\cdots$} }
\startdata
\\
\multicolumn{1}{c}{$0.3 < z < 0.6$}\\
\input{tab4a-data.tex}
\\
\multicolumn{1}{c}{$0.6 < z < 0.8$}\\
\input{tab4b-data.tex}
\\
\enddata
\tablenotetext{\dag}{$L_{\rm X}$ is quoted in the 0.5--2.0 keV band}

\end{deluxetable*}

%% file: tab4a-data.tex
 0.355 &  0.298 &  0.422 & 9.52 $\times$ 10$^{-7}$ & 1.65 $\times$ 10$^{-7}$ & 3.14 $\times$ 10$^{-6}$ & 0.329 &  1 \\
 0.502 &  0.422 &  0.597 & 9.51 $\times$ 10$^{-7}$ & 4.97 $\times$ 10$^{-7}$ & 1.70 $\times$ 10$^{-6}$ & 0.351 &  4 \\
 0.710 &  0.597 &  0.844 & 1.04 $\times$ 10$^{-6}$ & 7.62 $\times$ 10$^{-7}$ & 1.40 $\times$ 10$^{-6}$ & 0.372 & 14 \\
 1.004 &  0.844 &  1.194 & 1.93 $\times$ 10$^{-7}$ & 1.21 $\times$ 10$^{-7}$ & 2.97 $\times$ 10$^{-7}$ & 0.429 &  7 \\
 1.420 &  1.194 &  1.688 & 1.20 $\times$ 10$^{-7}$ & 8.22 $\times$ 10$^{-8}$ & 1.71 $\times$ 10$^{-7}$ & 0.439 & 10 \\
 2.008 &  1.688 &  2.388 & 5.99 $\times$ 10$^{-8}$ & 4.12 $\times$ 10$^{-8}$ & 8.56 $\times$ 10$^{-8}$ & 0.440 & 10 \\
 2.840 &  2.388 &  3.377 & 3.05 $\times$ 10$^{-8}$ & 2.05 $\times$ 10$^{-8}$ & 4.45 $\times$ 10$^{-8}$ & 0.509 &  9 \\
 4.016 &  3.377 &  4.776 & 1.47 $\times$ 10$^{-8}$ & 9.26 $\times$ 10$^{-9}$ & 2.27 $\times$ 10$^{-8}$ & 0.507 &  7 \\
 5.679 &  4.776 &  6.754 & 1.39 $\times$ 10$^{-9}$ & 2.41 $\times$ 10$^{-10}$ & 4.59 $\times$ 10$^{-9}$ & 0.516 &  1 \\
 8.032 &  6.754 &  9.552 & 1.75 $\times$ 10$^{-9}$ & $\cdots$ & 1.75 $\times$ 10$^{-9}$ & $\cdots$ &  0 \\

%% file: tab4b-data.tex
 1.420 &  1.194 &  1.688 & 1.59 $\times$ 10$^{-7}$ & $\cdots$ & 1.59 $\times$ 10$^{-7}$ & $\cdots$ &  0 \\
 2.008 &  1.688 &  2.388 & 3.75 $\times$ 10$^{-8}$ & $\cdots$ & 3.75 $\times$ 10$^{-8}$ & $\cdots$ &  0 \\
 2.840 &  2.388 &  3.377 & 6.76 $\times$ 10$^{-9}$ & 1.17 $\times$ 10$^{-9}$ & 2.23 $\times$ 10$^{-8}$ & 0.699 &  1 \\
 4.016 &  3.377 &  4.776 & 2.90 $\times$ 10$^{-9}$ & 5.02 $\times$ 10$^{-10}$ & 9.57 $\times$ 10$^{-9}$ & 0.625 &  1 \\
 5.679 &  4.776 &  6.754 & 1.51 $\times$ 10$^{-9}$ & 2.61 $\times$ 10$^{-10}$ & 4.98 $\times$ 10$^{-9}$ & 0.700 &  1 \\
 8.032 &  6.754 &  9.552 & 1.65 $\times$ 10$^{-9}$ & $\cdots$ & 1.65 $\times$ 10$^{-9}$ & $\cdots$ &  0 \\

%% file: tab5.tex
\begin{deluxetable*}{ccccccccccc}
\tablewidth{0pt} 
\tablecolumns{10} 
\tablecaption{Observed versus Expected Number of Clusters ($L_{\rm X, min} < L_{\rm X} < \infty$) \label{tab:counts}}

\tablehead{\colhead{$L_{\rm X, min}$\tablenotemark{\dag}} &
           \colhead{ $N_{\rm obs}$} & 
	   \colhead{} &
           \multicolumn{3}{c}{$N_{\rm exp}$} &
	   \colhead{} &
           \multicolumn{3}{c}{significance ($\sigma$)} \\
	   \noalign{\vspace{4pt}} \cline{4-6} \cline{8-10} \noalign{\vspace{0pt}} \colhead{($10^{44}$ $h_{50}^{-2}$ ergs s$^{-1}$)}\\
	   \colhead{} &
           \colhead{} &
	   \colhead{} &
           \colhead{REFLEX} &
           \colhead{BCS} &
           \colhead{RASS1BS} &  \colhead{} &
           \colhead{REFLEX} &
	   \colhead{BCS} &
           \colhead{RASS1BS}}
\startdata 
\\
\multicolumn{4}{l}{$0.3 < z < 0.6$, EdS} \\
\input{tab5a-data.tex}         
\\
\multicolumn{4}{l}{$0.3 < z < 0.6$, $\Omega_{M}=0.3$ and $\Omega_{\Lambda} = 0.7$} \\
\input{tab5b-data.tex}   
\\ 
\multicolumn{4}{l}{$0.6 < z < 0.8$, EdS} \\
\input{tab5c-data.tex}         
\\   
\multicolumn{4}{l}{$0.6 < z < 0.8$, $\Omega_{M}=0.3$ and $\Omega_{\Lambda} = 0.7$ } \\
\input{tab5d-data.tex}         
%\\   
\enddata
\tablenotetext{\dag}{Values of $L_{\rm X, min}$ (0.5--2.0 keV) are based on the lower limits of the luminosity bins used in the derivation of the non-parametric XLF}
\end{deluxetable*}

%% file: tab5a-data.tex
4.776&\phantom{0}0&\phantom{0}&\phantom{0}4.5&\phantom{0}3.4&\phantom{0}5.2&\phantom{0}&2.3&1.8&2.6\\
3.377&\phantom{0}2&\phantom{0}&\phantom{0}9.1&\phantom{0}6.7&10.8&\phantom{0}&2.5&1.8&3.0\\
2.388&\phantom{0}7&\phantom{0}&16.2&11.8&19.2&\phantom{0}&2.4&1.3&3.0\\
1.688&17&\phantom{0}&25.8&18.7&30.4&\phantom{0}&1.7&0.2&2.5\\

%% file: tab5b-data.tex
6.754&\phantom{0}0&\phantom{0}&\phantom{0}3.5&\phantom{0}3.0&\phantom{0}4.0&\phantom{0}&1.9&1.6&2.1\\
4.776&\phantom{0}1&\phantom{0}&\phantom{0}7.8&\phantom{0}6.1&\phantom{0}9.1&\phantom{0}&2.7&2.2&3.1\\
3.377&\phantom{0}8&\phantom{0}&14.8&11.0&17.4&\phantom{0}&1.7&0.7&2.3\\
2.388&17&\phantom{0}&24.4&17.8&28.8&\phantom{0}&1.4&0.0&2.2\\

%% file: tab5c-data.tex
4.776&\phantom{0}0&\phantom{0}&\phantom{0}4.1&\phantom{0}3.2&\phantom{0}4.8&\phantom{0}&2.1&1.7&2.4\\
3.377&\phantom{0}1&\phantom{0}&\phantom{0}7.9&\phantom{0}5.8&\phantom{0}9.3&\phantom{0}&2.7&2.0&3.1\\
2.388&\phantom{0}1&\phantom{0}&12.6&\phantom{0}9.1&14.9&\phantom{0}&3.9&3.1&4.4\\
1.688&\phantom{0}2&\phantom{0}&17.2&12.5&20.2&\phantom{0}&4.4&3.4&5.0\\
1.194&\phantom{0}3&\phantom{0}&20.5&15.0&24.1&\phantom{0}&4.6&3.5&5.2\\
0.844&\phantom{0}3&\phantom{0}&22.0&16.1&25.7&\phantom{0}&4.9&3.8&5.5\\

%% file: tab5d-data.tex
6.754&\phantom{0}0&\phantom{0}&\phantom{0}3.8&\phantom{0}3.3&\phantom{0}4.4&\phantom{0}&2.0&1.8&2.2\\
4.776&\phantom{0}1&\phantom{0}&\phantom{0}7.8&\phantom{0}6.1&\phantom{0}9.0&\phantom{0}&2.7&2.2&3.0\\
3.377&\phantom{0}2&\phantom{0}&12.7&\phantom{0}9.6&14.9&\phantom{0}&3.4&2.7&3.9\\
2.388&\phantom{0}3&\phantom{0}&17.4&12.9&20.5&\phantom{0}&4.0&3.1&4.6\\
1.688&\phantom{0}3&\phantom{0}&20.6&15.2&24.2&\phantom{0}&4.6&3.6&5.2\\
1.194&\phantom{0}3&\phantom{0}&21.8&16.1&25.6&\phantom{0}&4.8&3.8&5.5\\